\documentclass[11pt,a4paper]{article}
 \usepackage{jheppub}
\usepackage{amsmath,amssymb}
\usepackage{dsfont}
\usepackage{wrapfig}
\newcommand{\be}{\begin{equation}}
\newcommand{\ee}{\end{equation}}
\newcommand{\Pp}{\mathrm{P}} 
\title{Impure  Aspects of Supersymmetric Wilson Loops}
\author[a]{Valentina Cardinali} 
\author[b]{Luca Griguolo} 
\author[a]{and Domenico Seminara}
\affiliation[a]{Dipartimento di Fisica e Astronomia, Universit\`a di
Firenze and INFN Sezione di Firenze,\\ Via  G. Sansone 1, 50019 Sesto Fiorentino, Italy} 
\affiliation[b]{Dipartimento di Fisica, Universit\`a di Parma and INFN Gruppo Collegato di Parma, \\ Viale G.P. Usberti 7/A, 43100 Parma, Italy} 
\emailAdd{cardinali@fi.infn.it} 
\emailAdd{luca.griguolo@fis.unipr.it} 
\emailAdd{seminara@fi.infn.it}
\abstract{We study a general class of supersymmetric Wilson loops operator in ${\cal N}=4$ super Yang-Mills theory, obtained as orbits of conformal transformations. These loops are the natural generalization of the familiar circular Wilson-Maldacena operator and their supersymmetric properties are encoded into a Killing spinor that is not $pure$. We present a systematic analysis of their scalar couplings and of the preserved supercharges, modulo the action of the global symmetry group, both in the compact and in the non-compact case. The quantum behavior of their expectation value is also addressed, in the simplest case of the Lissajous contours: explicit computations at weak-coupling, through Feynman diagrams expansion, and at strong-coupling, by means of AdS/CFT correspondence, suggest the possibility of an exact evaluation.}
\begin{document} 
\maketitle

\section{Introduction}
Loop operators are probably the most basic observables of four dimensional gauge theories: they
can be classified according to whether the particle running around the loop is electrically or magnetically
charged, giving rise to Wilson or 't Hooft operators respectively. They play the role of order parameters for the phases
that a gauge theory can exhibit, and serve as probes of the quantum gauge dynamics. In supersymmetric gauge theories
loop operators become also ideal probes for checking some powerful, nonperturbative symmetry, as S-duality, that is conjectured to
exchange weak and strong coupling behaviors. The possibility to compute exactly these observables allows for a quantitative study
of S-duality and serves as a theoretical laboratory for gaining a deeper understanding of the inner workings of dualities among theories in 
different dimensions \cite{Gaiotto:2009we,Alday:2009aq,Alday:2009fs,Drukker:2009id}. On the other hand,
exact results in quantum field theory usually rely on powerful symmetry principles, such as supersymmetry: we could expect that particular loop operators, preserving some genuine fraction of the original supersymmetric invariance, are amenable of an exact quantum evaluation. A beautiful example is Pestun's calculation of circular 1/2 BPS Wilson-Maldacena loops \cite{Maldacena:1998im,Rey:1998ik} in a wide class of ${\cal N} = 2$ supersymmetric Yang-Mills theories \cite{Pestun:2007rz}. To be more
precise, Pestun reduced the problem of computing this highly supersymmetric observable to a finite-dimensional matrix integral, proving and generalizing the statement of a conjecture, originally formulated in the ${\cal N}=4$ case \cite{Er,Drukker:2000rr}. It appears therefore important to find new BPS loop operators and to study these non-local gauge invariant observables at quantum level.

We consider Wilson loops in four-dimensional maximally supersymmetric Yang-Mills theory in Euclidean space-time: ${\cal N} = 4$ SYM is a superconformal theory, the fermionic subspace of its superconformal algebra being generated by Poincar\'e supercharges $Q_\alpha$ and special conformal supercharges $S^\alpha$. We call a Wilson loop supersymmetric if there exists at least one non-zero linear combination of $Q_\alpha$ and $S^\alpha$ leaving invariant the operator and we are interested in  observables obtained from the ordinary electric loops by coupling them to the scalars of the ${\cal N} = 4$ supermultiplet. A certain number of such supersymmetric Wilson loops have been known for some time and analyzed previously \cite{Er,Drukker:2000rr,Zarembo:2002an,Drukker:2007qr}. They were captured by two classes: the loops of arbitrary shape found by Zarembo in \cite{Zarembo:2002an} and the loops of arbitrary shape on a three-sphere $S^3$, embedded into space-time, found by Drukker-Giombi-Ricci-Trancanelli (DGRT) in \cite{Drukker:2007qr}. Remarkably Zarembo's observables are the same Wilson loops which appear in topological Langlands twist of ${\cal N} = 4$ SYM \cite{Kapustin:2006pk} and have trivial expectation value. The most familiar example of the loops in DGRT class is instead the 1/2 BPS circular loop coupled to one of the scalars: it can be computed exactly by Gaussian matrix model and the results perfectly agree with the string dual computation, suggested by AdS/CFT correspondence. The subset of DGRT loops restricted to $S^2$ was also recently studied in great details and an interesting connection with bosonic two-dimensional Yang-Mills on $S^2$ was established \cite{Drukker:2007qr,Young:2008ed,Bassetto:2008yf,Giombi:2009ms,Bassetto:2009rt,Bassetto:2009ms,Giombi:2009ds,Pestun:2009nn}.

An essential step in understanding the structure of supersymmetric Wilson loops in ${\cal N} = 4$ SYM was performed by Dymarsky and Pestun in \cite{Dymarsky:2009si}: they were able to list all possible Wilson operators $\mathcal{W}$ that are invariant  under at least one superconformal symmetry $Q$ and to classify the interesting subclasses of $\{\mathcal{W},Q\}$ pairs modulo 
 the action of the superconformal group. The main idea in their construction is to pack the data describing locally a supersymmetric Wilson loop, namely the tangent vector to the curve and the scalar couplings, into a ten-dimensional vector $v^{M}(x)$. Requiring the invariance of the loop operator with respect to a supersymmetry $Q_\epsilon$ generated by a given spinor $\epsilon(x)$ implies a certain system of linear equations on $v(x)$. The properties of this system depend crucially on
whether the ten-dimensional spinor is {\it  pure} or not. Actually the appearance of pure spinors is not completely surprising because the four-dimensional theory is a dimensional reduction \cite{Brink:1976bc} of the ${\cal N} = 1$ SYM  in ten dimensions, where pure spinors show up naturally \cite{Nilsson:1985cm,Howe:1991mf,Howe:1991bx}. We remark that the space-time dependent  spinor that parameterizes the superconformal transformations of ${\cal N} = 4$ SYM, can be viewed directly as a reduction of a chiral ten-dimensional spinor. If $\epsilon(x)$ is not a {\it pure} spinor, then the system for $v(x)$ has the unique solution, {\it i.e.} the tangent to the curve and the scalar couplings are completely fixed. The vector $v(x)$ is is determined by the ten-dimensional vector constructed in the canonical way as the bilinear in $\epsilon(x)$. The contours obtained in this way from a general supersymmetry parameter $\epsilon(x)$ are simply the orbits of the conformal transformation generated by $Q^2_\epsilon$ \cite{Dymarsky:2009si}. Interestingly, modulo conformal equivalence, the only resulting compact curves are
the $(p,q)$ Lissajous figures where $p/q$ is the rational ratio of two eigenvalues of the $SO(4)$ matrix representing the action of $Q^2$. The situation changes if $\epsilon(x)$ is {\it pure}: in this case there are more solutions for the vector $v(x)$. Dymarsky and Pestun observed that a pure spinor defines ten-dimensional almost complex structure $J(x)$, and then the supersymmetry condition of the Wilson loop translates into the
condition that $v(x)$ is anti-holomorphic vector with respect to $J(x)$. On the subspace of the space-time where $\epsilon(x)$ is pure there is richer space of solutions for supersymmetric Wilson operator: generically, for any curve sitting inside the subspace one can find scalar couplings
to make  the Wilson loop supersymmetric.

The analysis presented in \cite{Dymarsky:2009si} is mainly concentrated on the classification and the classical construction of pure spinor Wilson loops, that are, in a sense, more general and interesting than the {\it impure} ones, being supported on rather arbitrary curves and admitting a strong-coupling characterization in type IIB superstring theory, as calibrated surface on $AdS_5\times S^5$. In this paper we are instead focussed on the less ambitious goal of studying  the 
{\it impure} Wilson loop operators and  their quantum aspects, both at weak and strong coupling. Our main concern is the compact case, therefore we study in details the $(p,q)$ Lissajous figures and their supersymmetry algebra: we find that, generically, five scalars couple to the supersymmetric Lissajous loop through a $6\times4$ constant rectangular matrix $M$ and a constant vector $B$. Both $M$ and $B$ generically possess complex entries and obey to some constraints that we solve explicitly. We recognize an apparent similarity with the scalar couplings introduced by Zarembo in \cite{Zarembo:2002an} except for the additional coupling governed by $B$. This small deformation plays a crucial role since it prevents the loops from having a trivial VEV, as occurs for the ones considered in \cite{Zarembo:2002an}. They also differ from the geometrically similar $toroidal$ $loops$, introduced by \cite{Drukker:2007qr}, where only three scalars are coupled. The loops are generically $1/16$ BPS but we observe enhancement of the supersymmetry for particular choices of the couplings. We also study the non-compact impure loops: we classify the orbits of the conformal group, writing down all the relevant contours modulo conformal equivalence and the corresponding couplings. We found convenient to rephrase this problem in six dimensional language, solving the orbit equations up to the action of an element of $SO(5,1)$: in so doing we construct some new families of supersymmetric loops, as logarithmic spirals, helix and generalized straight lines. At quantum level we consider specifically the Lissajous Wilson loops: at weak coupling, the most striking property is that the combined vector-scalar propagator in Feynman gauge, stretching on the Wilson loop contour, is constant, exactly as for the circular case. Moreover an explicit two-loops evaluation shows that the contribution of interacting diagrams to the quantum expectation value sums to zero, suggesting that the exact answer could be obtained by summing only exchanged propagators on the loop contour: if this would be the case, a Gaussian matrix model underlies the computation and an exact localization procedure should be invoked. We test directly this possibility at strong coupling, by using the dual description of Lissajous Wilson loops by strings in $AdS_5\times S^5$. More precisely the string duals propagate on a complexification of this space, as pointed out earlier in \cite{arXiv:0902.4586} where some cases of supersymmetric Wilson loops with complex scalar couplings were studied: we find indeed a perfect agreement with a Gaussian matrix model behavior. 

The main question opened to future investigations is, of course, if localization could provide an exact computation of this class of supersymmetric Wilson loops, reproducing the weak and strong coupling results we have found in this paper: it should certainly rely on some generalization of Pestun's procedure. The magnetic duals should also be constructed and studied at quantum level, providing new tests of S-duality. One could also wonder if some of the non-compact loops we found could be used in describing, at dual level, scattering processes or other observables in ${\cal N}=4$ SYM, for example constructing generalized cusps
\cite{arXiv:1105.5144}:  helixes, similar to the ones appearing in this paper, have also been considered in \cite{arXiv:0908.3020}.

The organization of the paper is the following: in Section 2 we briefly review the general strategy to classify supersymmetric Wilson loops in maximally supersymmetric four-dimensional Yang-Mills theory. In Section 3 we study Lissajous supersymmetric Wilson loops: we construct explicitly the scalar couplings and discuss their BPS properties and supersymmetry algebra. In Section 4 we perform the weak coupling computation at the second order in perturbation theory. In Section 5 we obtain the strong coupling solution by means of AdS/CFT correspondence. A number of Appendices is devoted to more technical aspects: Appendix \ref{app1} contains our conventions, Appendix \ref{Aapp} is dedicated to the complete classification of the conformal orbits, to the classification of scalar couplings and to the explicit construction of the relevant Killing spinors. In Appendix \ref{app3}  the action of  conformal transformations on the loops are discussed.

  \section{Impure Wilson loops}
In $\mathcal{N}=4$ super Yang-Mills (SYM)\footnote{See appendix \ref{app1} for our conventions on its action.}
the most simple and common generalization of  the familiar  Wilson loop is obtained by considering  extra couplings with the adjoint scalars $\Phi_{a}$ ($a=1,\dots,6$) \cite{Rey:1998ik,Maldacena:1998im}, namely by writing
\begin{equation}
\label{WL}
\mathcal{W}_R(\gamma)=\frac{1}{d(R)}\mathrm{Tr}_R \left[\Pp\!\exp\oint_\gamma \biggl(A_\mu (x(s)) \dot{x}^\mu(s)+\Phi_a  (x(s))  v^a(s)\biggr) ds\right]
\end{equation}
the suffix $R$ denoting the representation\footnote{The generator are taken anti-hermitian: $T^{A\dagger}=-T^{A}$.} of the  gauge group $G$ where the trace is  taken and  $d(R)$  its dimensions.  The 
six dimensional vector $v^{a}(s)$ identifies the new scalar couplings and, in general, its entries might  be complex in the euclidean case. For these operators  it is natural  to use a ten dimensional  notation (see  app.
\ref{app1}). In fact we can   combine the gauge field $A_{\mu}$ and the scalar fields $\Phi_{a}$ into a vector $A_{M}\equiv(A_{\mu},\Phi_{a})$ $(M=1,2,\dots,10)$ and we can merge  the tangent vector $\dot{x}^{\mu}$  and $v^{a}(s) $ into a  generalized  vector of couplings $v^{M}\equiv (\dot{x^{\mu}},v^{a})$. Then the Wilson loop \eqref{WL}
can be rearranged in the compact form \cite{Dymarsky:2009si}
\be
\label{WL1}
\mathcal{W}_R(\gamma)
=\frac{1}{d(R)}
\mathrm{Tr}_R\left[ \Pp\!\exp\left(\oint_\gamma A_Mv^M ds\right)\right].
\ee
An interesting subclass of these  non-local  operators is provided by the so-called {\it supersymmetric} Wilson loops, {\it i.e.} the operators \eqref{WL1}  for which the combination $v^{M} A_{M}$ is  invariant under, at least,  one super-conformal  transformation. This subset is determined by the vectors $v^{M}$ 
which obey  the linear constraint
\begin{equation}
\label{susycond}
\delta_{\epsilon}(A_{M} v^{M})= v^M(s)\psi\gamma_M\epsilon\left(x\right)=0\ \ \ \ \Rightarrow\ \ \  \
  v^M(s)\gamma_M\epsilon\left(x\right)=0,
\end{equation} 
where $\epsilon(x)=\epsilon_{s}+x^{\mu}\gamma_{\mu}\epsilon_{c}$ is the super-conformal Killing spinor associated to the transformation.
Locally on the contour this implies that $v^M v_M=0$ \cite{Drukker:1999zq}. More generally, given $\epsilon(x)$, all possible solutions for $v^{M}$  of eq. \eqref{susycond} were obtained by Dymarsky and Pestun in  \cite{Dymarsky:2009si}. They   fall into two different  classes depending on the value of the bilinear 
$u^{M}\equiv\epsilon^{T} C^{-1}\gamma^{M}\epsilon$.

\paragraph{\sc Case (A):} If  $u^{M}$ vanishes  identically
  on a submanifold $\Sigma_{\epsilon} \subseteq \mathds{R}^{4}$, $\epsilon$
  is a \textit{pure spinor} on $\Sigma_{\epsilon}$ and consequently it induces an almost complex structure $J_{\epsilon}$ on this region
  \cite{Dymarsky:2009si}.    The possible  solutions  $v^{M}$ of eq. \eqref{susycond} in a point $x\in \Sigma_{\epsilon}$  are   then provided by all the anti-holomorphic  vectors with respect to $J_{\epsilon}$   \cite{Dymarsky:2009si}.  This result can be used to associate a supersymmetric Wilson loop to each  closed contour $\gamma$ in $\Sigma_{\epsilon}$. An explicit  construction   of this class  of  operators, modulo equivalence under the action of the superconformal group, is given in \cite{Dymarsky:2009si}. All supersymmetric Wilson loops that have been studied previously are essentially captured by this case. In fact
 both  the loops discussed by Zarembo in \cite{Zarembo:2002an} and those found by Drukker-Giombi- Ricci-Trancanelli (DGRT) in \cite{Drukker:2007qr} are of this type.

\paragraph{\sc  Case (B):} When $u^M\neq 0$, the  solution of eq. \eqref{susycond} is  uniquely fixed up to a complex scale  $\lambda$ and it is given by $v^M=\lambda u^M$ \cite{Dymarsky:2009si}. In other words, given the super-conformal  spinor $\epsilon$,  there  is only one possible invariant Wilson loop:
\begin{equation}
\label{WLL}
\mathcal{W}_R\left({\gamma}\right)=\frac{1}{d(R)}\mathrm{Tr}_R\left[ \mathrm{P}\!\exp\left(\oint_{\gamma} \left(A_\mu u^\mu+\Phi_a u^a\right)\frac{ds}{\left(u^\mu u_\mu\right)^{1/2}}\right)\right],
\end{equation}
where  $s$ denotes the usual affine parameter which measures the length of the curve. In eq. \eqref{WLL}, in order to identify the space-time couplings $u^{\mu}$ with the tangent vector to the  contour $\gamma$, we must require that $u^{\mu}$ is projectively equivalent to a real vector, {\it i.e.} there is  a $\lambda\in\mathds{C}^{*}$ such that $\lambda u^{\mu}$ is real\footnote{The reality of $\lambda u^{\mu}$ is an implicit constraint on the possible spinors $\epsilon$.}. Then $\gamma$ is determined by the differential equation 
\be
\label{gamma}
\dot{x}^{\mu}= u^{\mu},
\ee
where,  for future convenience, we have chosen to fix the normalization of the spinor $\epsilon$ so that  
$\lambda=1$. The path $\gamma$ defined  by \eqref{gamma}  has a natural and simple   geometrical interpretation:  it  is  just  the orbit  of the conformal transformation generated by $Q^2_\epsilon$, where 
$Q_{\epsilon}$ is the superconformal  generator associated to the spinor $\epsilon$ (see \cite{Dymarsky:2009si}).

\medskip
\noindent
In the  following we shall focus our attention on the supersymmetric Wilson loops  of this  second type 
for which we  shall also use the term {\it impure loops} to emphasize the difference  with those of the 
{\sc Case (A)}. Specifically, we shall  determine the general form of the scalar couplings and provide an explicit construction of  this  family of loops. Next we shall discuss the complete sub-algebra  of the superconformal group which leaves these loop invariant. Finally we shall analyze the properties of their VEVs both 
at  weak and at strong coupling. 

\section{ Lissajous figures: general properties and structure}
\label{Genstruct}
In general, the integral curves of the  conformal Killing vector $u^{\mu}$ do not define a closed loop: it can only occur when $u^{\mu}$, modulo conformal equivalence,  specifies an orbit of the four dimensional rotation group. 
 In this case $u^{\mu}$ can be
always cast into the canonical form 
\be
\label{umu}
u^{\mu}=\Omega^{\mu}_{\,\, \nu} x^{\nu},
\ee
  (see app. \ref{vuop}.) The matrix $\Omega$ in \eqref{umu}  can be chosen to belong to  the (Cartan) subalgebra $so(2)\oplus so(2)$ in $so(4)$ and its explicit  form is
\be
\label{omegatilde}
 \Omega =\begin{pmatrix} 0 &\Omega_{1} & 0 &0\\ -\Omega_{1} & 0 & 0 & 0\\ 0 & 0 & 0 & \Omega_{2}\\ 0 & 0 &-\Omega_{2} & 0
\end{pmatrix}.
\ee
Then the orbits drawn by the tangent vector \eqref{umu}  are  described by the parametric equations\footnote{We have dropped an irrelevant global scale in solving  \eqref{gamma} and used the freedom to choose the initial point of the loop.}
\be
\label{liss}
x^{\mu}(s)=\left\{\cos \frac{\theta}{2}\sin\Omega_{1} s,\ \cos \frac{\theta}{2}\cos\Omega_{1} s,\ \sin \frac{\theta}{2}\sin \Omega_{2} s,\ \sin \frac{\theta}{2}\cos \Omega_{2} s\right\},
\ee
where $\theta$ is a free parameter which runs from $0$ to $\pi$. However, these paths define a closed circuit 
 if and only if  the ratio $\Omega_{2}/\Omega_{1}$ is a rational number  $\displaystyle{m/n}$ with $m,n$  relatively prime. In this case the range of $s$ must be a multiple of $\displaystyle{\frac{2\pi n}{\Omega_{1}}} $. 

\begin{figure}[ht]
\centering
	\includegraphics[width=.3 \textwidth]{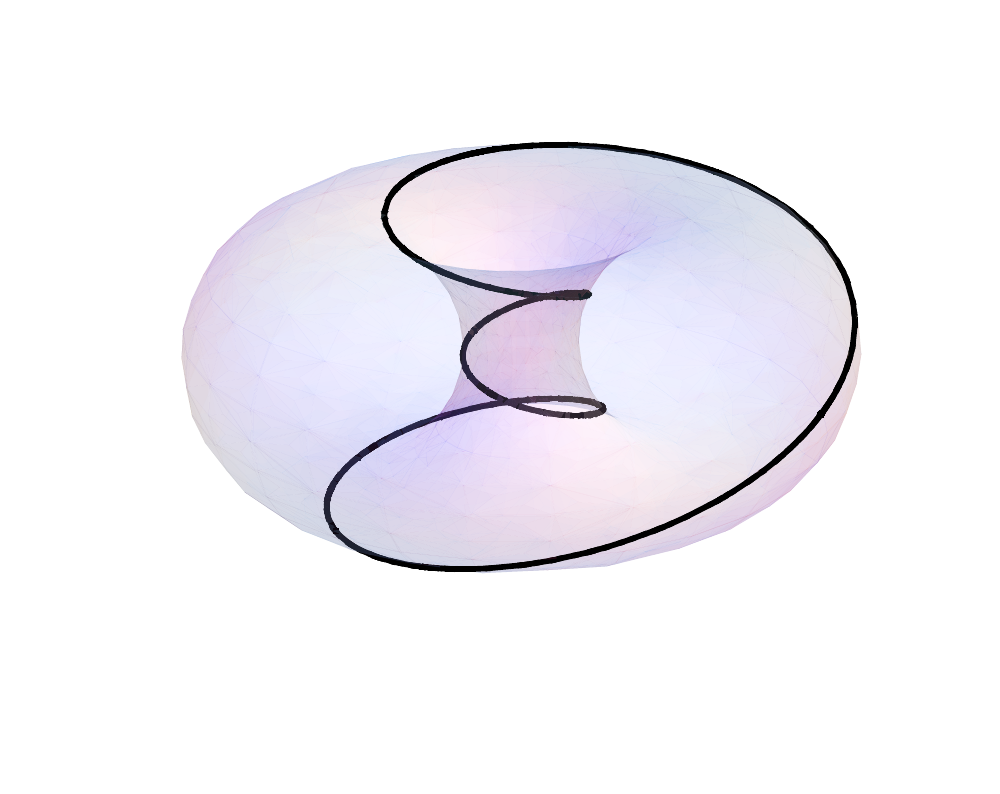}
	\includegraphics[width=.3 \textwidth]{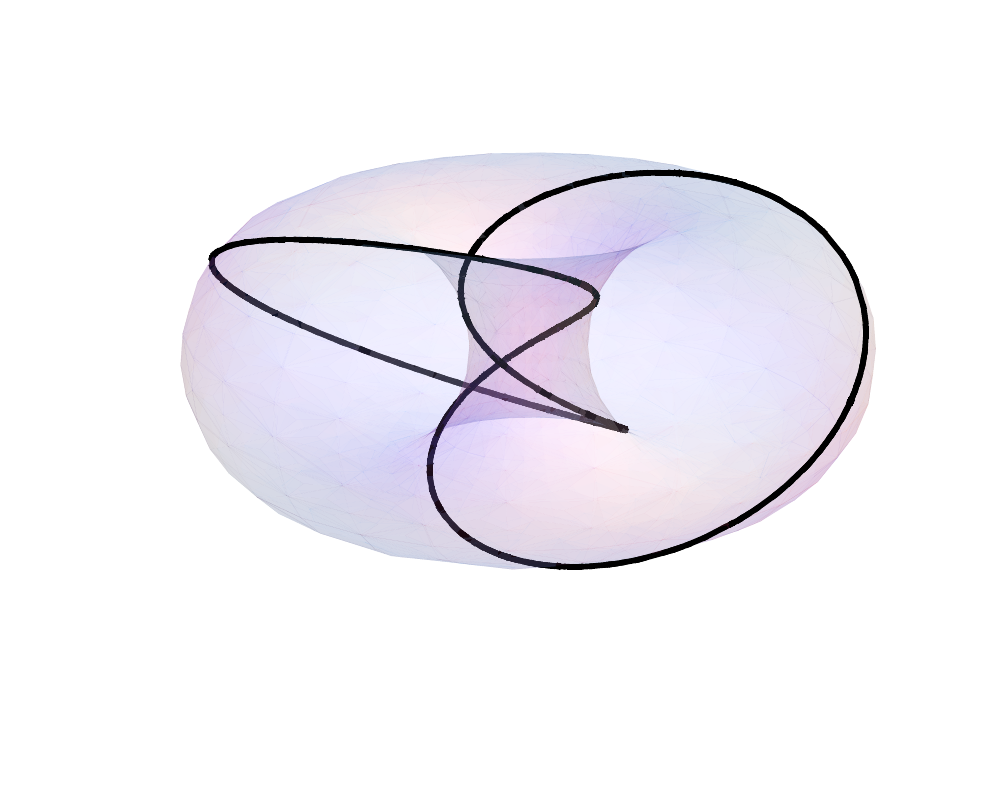}
	\put(-199,5){(a)}	\put(-62,5){(b)}
\caption{\label{torus}
For fixed $\theta$ all these loops wrap a torus  $T^{2}$ of equation $\left(\sqrt{y_1^2+y_2^2}-\sec \frac{\theta }{2}\right)^2+y_3^2=\tan ^2\frac{\theta }{2}$.  Fixing $\theta=\pi/2$,  the  loops for $\Omega_{2}/\Omega_{1}=1/3$ and $\Omega_{2}/\Omega_{1}=2/3$ are shown in (a) and (b) respectively.}
\end{figure}

\noindent
Geometrically, the curves \eqref{liss} describe the superposition of two circular motions  with different frequencies occurring 
in orthogonal planes and they  can be considered a generalization  of the familiar   {\it Lissajous 
figures}.  By construction they all lie on the sphere $S^{3}$ defined by $x_{1}^{2}+x_{2}^{2}+x_{3}^{2}+x_{4}^{2}=1$ and thus we can use the  usual stereographic coordinates  $\left (y_{i}\equiv\frac{x_{i}}{1-x_{4}} \ \  i=1,2,3\right)$ to picture them (see Fig. \ref{torus}).

When the ratio   $\Omega_{2}/\Omega_{1}$ runs from $0$ to $1$,  at fixed $\theta$, we are interpolating  between a latitude ($\Omega_{2}/\Omega_{1}=0$, the red curve in Fig.\ref{q=0,1})
on the  $S^{2}$ defined by $x_{3}=0$ and one of the equators of the sphere $S^{3}$ (the blue curve in Fig.\ref{q=0,1}). Instead, at fixed $\Omega_{2}/\Omega_{1}$, 
when the parameter $\theta$ goes from $0$ to $\pi$, we have a family of circuits  interpolating between two great circles  in $S^{3}$:
the former (at $\theta=0$) with  winding  number $n$, the latter  (at $\theta=\pi$) with winding  number $m$.\\
\begin{wrapfigure}[14]{l}{68mm}
\centering
    \includegraphics[width=.35 \textwidth, height=.25 \textwidth]{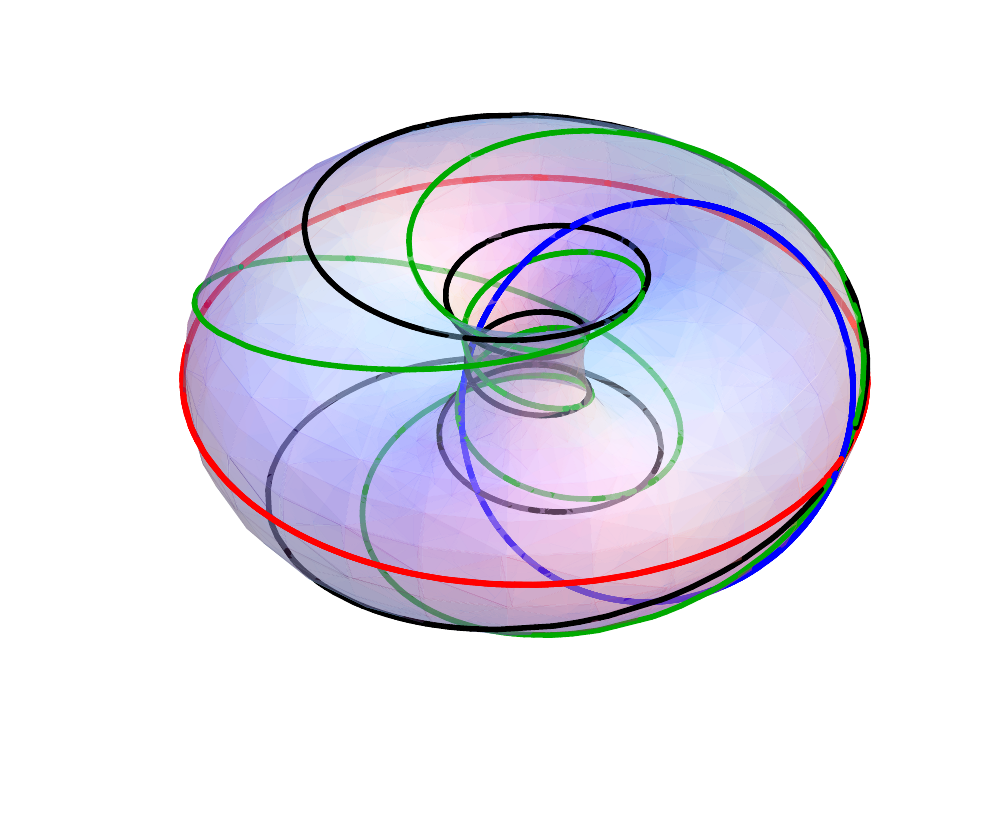}
\vskip -.3cm
\caption{\label{q=0,1}
The red curve corresponds to the latitude, while the blue one is a great circle of $S^{3}$.The black and  green contours corresponds to intermediate values of $\Omega_{2}/\Omega_{1}$.}
\end{wrapfigure}
\vskip -1.2cm
The same contours were considered  in \cite{Drukker:2007qr} as an example of DGRT  loops with more than 2 super-symmetries and they were named {\it toroidal loops}. However, as we shall see below,  the scalar  couplings are quite different in the two cases. For example, while  DGRT  loops in general couple to three 
scalars, ours will  couple  to five. Additional differences will become manifest when discussing the structure of the scalar couplings below.

For this family of operators the vector $u^{a}$  $(a=1,\dots,6)$, which couples the contour to the scalar fields $\Phi_{a}$,  has a very simple structure. Up to terms vanishing along the loop,  it is linear in ${x}^{\mu}$ or alternatively in $\dot{x^{\mu}}$ and it can be written as follows  
\be
\label{coupl1a}
 u^{a}= M^{a}_{\,\, \mu}x^{\mu}+B^{a}=(M\Omega^{-1})^{a}_{\ \ \mu}\dot{x}^{\mu}+B^{a},
\ee
where $M$ is a $6\times 4$ constant rectangular matrix   and $B$ is a constant vector. Both $M$ and $B$ generically possess complex entries. In addition  the matrix  $M$ must obey the 
following algebraic  constraints
\be
\label{localsusy}
(B\cdot M)_{\mu}=0\ \  \  \ \  ^{t}\!M M+ ^{t}\!\Omega\Omega=-(B\cdot B)\mathds{1}_{4\times 4},
\ee
which encode the requirement  of local supersymmetry $[{\it i.e.}\ \ u^{M} u_{M}=0]$. 

\noindent
The {\it impure} loops will commonly couple to five  independent scalars. In fact, because of eqs. \eqref{localsusy}, the columns of $M$ and the vector $B$  will provide  in general a set 
of 5 orthogonal and thus independent vectors\footnote{Keep also in mind that the five functions $\{1,x_{\mu}\}$ are linear independent for generic  $\Omega_{1}$ and $\Omega_{2}$.}. This is a crucial difference with the {\it toroidal loops}
considered in \cite{Drukker:2007qr} were the number of the coupled scalar was at most three.

In \eqref{coupl1a}  we  also recognize an apparent  similarity with the scalar couplings introduced  by Zarembo in \cite{Zarembo:2002an} except for the  additional coupling governed by $B^{A}$.  This small  deformation plays a crucial role since it  prevents the loops defined by \eqref{coupl1a} from having a trivial VEV, as occurs for the ones considered in \cite{Zarembo:2002an}. Actually,
from the point of view of perturbation theory, this family of loops can be considered the simplest generalization of the usual  Wilson-Maldacena circle and in fact it  enjoys very similar properties (see also Sect.\ref{Perb aspects}).

The details of the  construction and the properties of these non-local operator starting from the {\it impure} Killing spinor $\epsilon$ are presented in appendix \ref{Aapp}. 

\subsection{Supersymmetries}
The  Wilson loops introduced in Sect.\ref{Genstruct} are, by construction, invariant under the  super-conformal transformation defined by the spinor $\epsilon=\epsilon_{s}+x^{\mu}\gamma_{\mu}\epsilon_{c}$, which generates the 
vector of the couplings.  
  But  this native  invariance does not exhaust all possible  supersymmetries. To classify all of them, we must solve the standard BPS-condition
\be
\label{figi0}
\delta ({u^{M} A_{M}})\propto (\dot{x}^{\mu}\gamma_{\mu}+ u^{a}\gamma_{a})\epsilon=
 (\dot{x}^{\mu}\gamma_{\mu}+ u^{a}\gamma_{a})(\epsilon_{s}+x^{\mu}\gamma_{\mu}\epsilon_{c})=0,
\ee
where  we used the usual $32\times 32$ Dirac matrices $\gamma_{M}$ to have a more efficient notation and we have also broken the range of their index $M$ in two subsets $(\mu,a)$ with $\mu=1,\dots,4$ and 
 $a=1,\dots,6$.
 
We can reorganize eq. \eqref{figi0} as a polynomial of second degree in the space-time coordinates $x^{\mu}$. Then it takes the following form
\be
\label{figi}
\begin{split}
&x^{\nu}[\sigma_{\nu}\epsilon_{s}+(B_{a}\gamma^{a})\gamma_{\nu}\epsilon_{c}]- (x\cdot\gamma)
x^{\nu}[ (B_{a}\gamma^{a})\gamma_{\nu}\epsilon_{s}+\sigma_{\nu}\epsilon_{c}]=0,
\end{split}
\ee
where we have  used that our closed contours must lie on the unit sphere,  namely they obey the constraint $x^{2}=1$. In eq. \eqref{figi} we have also found convenient to introduce the auxiliary  matrices
\be
\sigma_{\alpha}=\left(\gamma_{\mu}\Omega^{\mu}_{\ \ \alpha} +\gamma_{a}
M^{a}_{\,\, \alpha}\right).
\ee
Because of eq. \eqref{localsusy} they  obey the four dimensional Clifford algebra
\be
\label{ww1}
\sigma_{\alpha}\sigma_{\beta}+\sigma_{\beta}\sigma_{\alpha}=-2(B\cdot B)\delta_{\alpha\beta}\mathds{1}\ \ \  \ \
\ee
and a very simple set of  anti-commutation relations with the standard Dirac matrices $\gamma_{\nu}$ and with
  $B_{a}\gamma^{a}$:
\be
\label{ww2}
\gamma^{\mu} \sigma_{\nu}+\sigma_{\nu} \gamma^{\mu}=2\Omega^{\mu}_{\ \ \nu}\mathds{1} \  \ \  \ \  \ {\rm and} \  \ \ \{\sigma_{\mu},B_{a}\gamma^{a}\}=0.
\ee
For a generic value of $\Omega_{1}$ and $\Omega_{2}$ [{\it i.e.} $\Omega_{1}^{2}\ne \Omega_{2}^{2}\ne 0$], the monomials  in $x^{\mu}$,
 appearing in 
\eqref{figi}, provide an independent set of functions along the circuit, 
thus the two combinations between square brackets must vanish separately
\begin{subequations}
\label{su}
\begin{align}
\label{su1}
&\sigma_{\nu}\epsilon_{s}+(B^{a}\gamma_{a})\gamma_{\nu}\epsilon_{c}=0, \\
\label{su1a}
& (B_{a}\gamma^{a})\gamma_{\nu}\epsilon_{s}+\sigma_{\nu}\epsilon_{c}=0.
\end{align}
\end{subequations}
For $(B\cdot B)\ne 0$, we can ignore the conditions \eqref{su1a}  since they can be shown to be  equivalent
to the set of equations \eqref{su1}. Thus we are left with only four equations constraining the couple of constant spinors
$(\epsilon_{s},\epsilon_{c})$. To solve this system, we  first get rid of  $\epsilon_{c}$ by solving  \eqref{su1} for $\nu=1$ 
\be
\label{pw}
\epsilon_{c}=\frac{1}{(B\cdot B)} (B^{a}\gamma_{a}) \gamma^{1}\sigma_{1}\epsilon_{s}.
\ee
Substituting \eqref{pw} into the remaining  three equations we learn that $\epsilon_{s}$ is annihilated   by the  following three linear operators
\begin{subequations}
\label{pw1}
\begin{align}
\label{T1eq1}
\mathds{T}_{1}\epsilon_{s}&\equiv (\gamma^{2}\sigma_{2}-\gamma^{1}\sigma_{1})\epsilon_{s}=
[\gamma^{2}\hat M_{2}-\gamma^{1}\hat M_{1}]\epsilon_{s}=0,\\
\mathds{T}_{2}\epsilon_{s}&\equiv(\gamma^{3}\sigma_{3}-\gamma^{2}\sigma_{2})\epsilon_{s}=
[(\Omega_{1}\gamma^{12}-\gamma^{2}\hat M_{2})-(\Omega_{2}\gamma^{34}
-\gamma^{3}\hat M_{3})]\epsilon_{s}=0,\\
\mathds{T}_{3}\epsilon_{s}&\equiv(\gamma^{4}\sigma_{4}-\gamma^{3}\sigma_{3})\epsilon_{s}=
[\gamma^{4}\hat M_{4} -
\gamma^{3}\hat M_{3}]\epsilon_{s}=0,
 \end{align}
\end{subequations}
where we have introduced the short-hand notation $\hat M_{\mu}\equiv \gamma_{a}M^{a}_{\ \ \mu}$. Since the six-component vectors $(M^{a}_{\ 1},
M^{a}_{\ 2}, M^{a}_{\ 3}, M^{a}_{\ 4})$ generically define four orthogonal complex directions in $\mathds{C}^{6}$ [see eq. \eqref{localsusy}], the projectors on the  kernels of  the  operators
$\mathds{T}_{i}$ are easily constructed in terms of the matrices $\gamma^{\mu}$ and $\hat M_{\mu}$.
One finds
\begin{align}
\label{fd}
\mathds{P}_{1}&=\frac{1}{2}\left(\mathds{1}-\frac{(\gamma^{2} \hat M_{ 2 })(\gamma^{1} \hat M_{ 1 })}{(B\cdot B)+\Omega_{1}^{2}}\right),\nonumber\\
\mathds{P}_{2}&=\frac{1}{2}\left(\mathds{1}-\frac{ (\Omega_{1}\gamma^{12}-\gamma^{2}\hat M_{2} )(\Omega_{2}\gamma^{34}
-\gamma^{3}\hat M_{3})}{( B\cdot B)}\right), \nonumber\\
\mathds{P}_{3}&=\frac{1}{2}\left(\mathds{1}-\frac{(\gamma^{4} \hat M_{4})(\gamma^{3} \hat M_{3})}{(B\cdot B)+\Omega_{2}^{2}}\right).
\end{align}

As they  commute  among themselves,  the most general solution of the \eqref{pw1}  can be always
cast  into the form
\be
\label{cito1}
\epsilon_{s}=\mathds{P}_{1} \mathds{P}_{2} \mathds{P}_{3}\mathds{\eta}_{s},
\ee
where $\eta_{s}$ is a positive chiral spinor in ten dimensions.  The number of  linearly independent solutions  can be also easily determined: in fact it is equal  to  the  rank of the projector $\mathds{P}_{1} \mathds{P}_{2} \mathds{P}_{3}$ on the subspace of spinors of positive chirality. This last quantity is simply obtained by taking the  trace of the combination $\frac{1}{2}(1+\gamma^{11})\mathds{P}_{1} \mathds{P}_{2} \mathds{P}_{3}$:
\be
\mathrm{Tr}\left(\frac{1}{2}(1+\gamma^{11})\mathds{P}_{1} \mathds{P}_{2} \mathds{P}_{3}\right)=2.
\ee
Our family of loops preserves generically two independent supercharges, being at least $1/16$ BPS. The degree of supersymmetry can be of course enhanced for particular choices of the scalar couplings. In particular we see that the above analysis, done in terms of commuting projectors, appears to fail when $(B\cdot B)$ is equal to either $-\Omega^{2}_{1}$  or $-\Omega_{2}^{2}$ or to $0$\footnote{Since $(B\cdot B)=-2 (B_{0}\cdot B_{1})$, these are  exactly the same singular points encountered in the general discussion of  the couplings in appendix \ref{vuop0}.}. For those values  the expression \eqref{fd} for one of the  three  projectors  is ill-defined. 

We start by considering the case $(B\cdot B)=-\Omega_{1}^{2}$ (but $\Omega_{1}^{2}\ne \Omega_{2}^{2}$)\footnote{The case $(B\cdot B)=-\Omega_{2}^{2}$ (but $\Omega_{1}^{2}\ne \Omega_{2}^{2}$)  can be analyzed in a similar way. It can be obtained from this one by exchanging the role of $\Omega_{1}$ and $\Omega_{2}$.}. It is not difficult to show from the constraints \eqref{localsusy} (see also Appendix B, above eq.\eqref{B32}) that the first two  columns of the matrix  $M^{a}_{\ \mu}$  are given by two complex parallel
{\it light-like} vectors: $M^{a}_{\ 1}=m_{1} V^{a}$,  $M^{a}_{\ 2}=m_{2} V^{a}$ and  $V^{2}=0$ . 
As a consequence, eq.\eqref{T1eq1} can be rearranged as follows
\be
\label{T1red}
\mathds{T}_{1}\epsilon_{s}=(m_{1}\gamma^{1}-m_{2}\gamma^{2})\hat V\epsilon_{s}=0,
\ee  
where we have defined $\hat V\equiv V^{a}\gamma_{a}$. If  $m_{1}^{2}\ne -m^{2}_{2}$, the kernel of $\mathds{T}_{1}$  is simply equivalent  to that of  $\hat V$ and the general solutions of our first equation can be written as $\hat V \eta$, being $\eta$ an arbitrary anti-chiral spinor. The solution of the full system (\ref{pw1}) is then obtained 
by applying the projector $\mathds{P}_{2}$ and $\mathds{P}_{3}$ on $\hat V \eta$,
\be
\label{df}
\epsilon_{s}=\mathds{P}_{2}\mathds{P}_{3}\hat V \eta,
\ee
and finding the independent components as $\eta$ is varied. The analysis can be performed in a pedestrian way and one ends up with only two independent spinor $\epsilon_{s_i}$, whose  explicit form is 
\be
\label{sold}
\epsilon_{s_i}=\sqrt{\Omega_{1}-\Omega_{2}}\epsilon_{s_i }^{+}+\sqrt{\Omega_{1}+\Omega_{2}}\epsilon_{s_i}^{-},
\ee
where $\epsilon_{s_i}^{\pm}$ are chiral spinors with respect to the matrix $\gamma^{1234}$
and they are given by
{\small
\begin{align}
\epsilon^{+}_{s_1}=&\left\{1,-i,0,0,0,0,0,0,0,0,0,0,0,0,0,0\right\}\nonumber\\
\epsilon^{+}_{s_2}= &\left\{4 p_{1}(B^1_{1}-i B^2_{1}),0,2 p_{1} (B^{3}_{1}-i B_{1}^{4}),2 i p_{1} (B_{1}^{3}-i B_{1}^{4}),0,0,0,0,0,0,0,0,-\Omega_{1}^2,i \Omega_{1}^2,0,0\right\}\nonumber\\ 
\epsilon^{-}_{s_1}=&\left\{0,0,0,0,i e^{i \alpha },e^{i \alpha },0,0,0,0,0,0,0,0,0,0\right\} \label{fgh}\\
\epsilon^{-}_{s_2}=&\left\{0,0,0,0,-4 p_{1} (B^1_{1}-i B^2_{1})\sin \alpha ,4 p_{1}  (B^1_{1}-i B^2_{1})\cos \alpha ,-2 i e^{-i \alpha } p_{1} (B^3_{1}+i B^4_{1}),\right.\nonumber\\
   &\left.-2 e^{-i \alpha }
   p_{1} (B^3_{1}+i B^4_{1}),-i e^{-i \alpha } \Omega_{1}^2,-e^{-i \alpha } \Omega_{1}^2,0,0,0,0,0,0
   \right\}\nonumber.
\end{align}}

\vskip -.7 cm
\noindent
The vector $B_{1}^{a}$ and the parameters  $p_{1}$ and $\alpha$ are defined in  appendices  \ref{GP} and \ref{vuop0}.  To avoid a cumbersome notation we have also dropped the last sixteen  entries, which obviously vanish for a spinor of positive chirality. 

Since we have only the two independent solutions \eqref{fgh},  the loops are still 1/16 BPS. However, in this particular case, they are not anymore {\it impure}: as one can easily check, any conformal Killing spinors $\epsilon$ associated to the solutions \eqref{sold} solves $\epsilon^{T}C^{-1}\gamma^{M}\epsilon=0$  on the unit sphere $S^{3}$. In other words they define a family of {\it pure} loops coupled to four scalars.

The next step is to explore the case $m^{2}_{1}+m^{2}_{2}=0$, that provides an enlarged space of solutions. Taking $m_{1}\ne 0$, besides the two conformal Killing spinors determined above, a third linearly independent solution surprisingly appears. For instance, for $m_{2}=i m_{1}$,  it is given by
\be
\epsilon_{s_3}=\sqrt{\Omega_{1}-\Omega_{2}}\epsilon_{s_3}^{+}+\sqrt{\Omega_{1}+\Omega_{2}}\epsilon_{s_3}^{-},
\ee
with
\be
{\small
\label{wer1}
\begin{split}
\!\!\!\!\epsilon^{+}_{s_3}=&\left\{-e^{i \alpha } \sqrt{\Omega _1-\Omega _2}-\frac{i m_1 \sqrt{\Omega _1+\Omega _2}}{\Omega _1}-\frac{e^{-i \alpha } }{8 \Omega _1 \sqrt{\Omega _1-\Omega _2}},-i e^{i \alpha }
   \sqrt{\Omega _1-\Omega _2}-\right.\\
&\left.-\frac{m_1 \sqrt{\Omega _1+\Omega _2}}{\Omega _1}+\frac{i e^{-i \alpha } }{8 \Omega _1 \sqrt{\Omega _1-\Omega _2}},\frac{m_4+i m_3}{\sqrt{\Omega
   _1+\Omega _2}},\frac{-m_3+i m_4}{\sqrt{\Omega _1+\Omega _2}},0,0,0,0,0,0,0,0,0,0,0,0\right\}
   \end{split}}
\ee
and
 \be
  {\small
  \label{wer2}
   \begin{split}
\epsilon^{-}_{s_3}=&   \left\{0,0,0,0,\frac{8 i \left(\Omega _1-\Omega _2\right) \left(\Omega _1 \sqrt{\Omega _1+\Omega _2}+i e^{i \alpha } m_1 \sqrt{\Omega _1-\Omega _2} \right)-i  \sqrt{\Omega _1+\Omega _2}}{8 \Omega _1 \sqrt{\Omega _1-\Omega _2} \sqrt{\Omega _1+\Omega _2}},\right.\\&\left.\frac{-\sqrt{\Omega _1+\Omega _2}\left(1+8 \Omega _1
   \left(\Omega _1-\Omega _2\right)\right)+8 i e^{i \alpha } m_1 \left(\Omega _1-\Omega _2\right){}^{3/2} }{8 \Omega _1 \sqrt{\Omega _1-\Omega _2}
   \sqrt{\Omega _1+\Omega _2}},\right.\\
& \left.  \frac{e^{i \alpha } \left(m_3+i m_4\right)}{\sqrt{\Omega _1+\Omega _2}},\frac{e^{i \alpha } \left(m_4-i m_3\right)}{\sqrt{\Omega _1+\Omega _2}},0,0,0,0,0,0,0,0\right\}.
\end{split}}
\ee
\noindent
The additional parameters $m_{i}$ in \eqref{wer1} and \eqref{wer2} are again defined in appendix \ref{vuop0}. The new spinor turns out to be {\it impure} and in fact the vector $\epsilon_{_3} C^{-1}\gamma^{M}\epsilon_{_3}$  is proportional to  the vector of couplings.

A further and more clear enhancement in the solutions space is observed for $m_{1}=m_{2}=0$, when the columns $M_{\ 1}^{a}$ and  $M^{a}_{\  2}$  do both   vanish and the operator $\mathds{T}_{1}$ is  identically zero. We loose eq.\eqref{T1eq1} and the most general solution of the remaining two equations can be readily written as 
\be
\epsilon_{s}=\mathds{P}_{2} \mathds{P}_{3}\mathds{\eta}_{s},
\ee
with  $\eta_{s}$ a spinor of positive chirality. We have an obvious augmentation of the supersymmetry and in fact the number of independent solutions of  \eqref{pw1} is now given by
\be
\mathrm{Tr}\left(\frac{1}{2} (1+\gamma^{11})\mathds{P}_{2} \mathds{P}_{3}\right)=4.
\ee
This particular subset of  loops is therefore  $1/8$ BPS and  the number of scalars coupled to the contour is conversely reduced from five to three. 

Finally, we consider the case $(B\cdot B)=0$ (with $\Omega_{1}^{2}\ne \Omega_{2}^{2}$ ), when the two sets of equations  \eqref{su} are not linearly dependent.  One can solve again both equations and verify that the loops are still $1/16$ BPS. 

\subsection{Supersymmetry algebra}
For a generic value of $\Omega_{1}$ and $\Omega_{2}$ [{\it i.e.} $\Omega_{1}^{2}\ne \Omega_{2}^{2}\ne 0$] and $(B\cdot B)$ different from either $-\Omega^{2}_{1}$  or $-\Omega_{2}^{2}$ or $0$, the {\it impure} loops are preserved by only two superconformal charges $Q_{i}$ ($i=1,2$), which are generated 
by the two independent solutions, $\epsilon_{i}=\epsilon_{s_i}+x^{\mu}\gamma_{\mu}\epsilon_{c_{i}}$ $(i=1,2)$, of the linear system \eqref{pw1}. In the following we shall analyze the  associated super-algebra.

To avoid a lengthy exercise in {\it spinorology}, we  focus our analysis on the little group preserving the 
origin of the coordinates $(x^{\mu}=0)$ \cite{Pestun:2009nn}. In this case for the anti-commutator of two super-charges can be
cast in the following form \cite{Pestun:2009nn}
\begin{equation}
\label{anticomm1}
Q_{\left\{i\right.}Q_{\left.j\right\}}=2 \epsilon_{{\left\{c_i\right.}}^{T}C^{-1}\gamma_{ab}\epsilon_{{\left. s_j\right\}}}R_{ab}-2\epsilon_{{\left\{c_i\right.}}^{T}C^{-1}\gamma_{\mu\nu}\epsilon_{{\left. s_j\right\}}}R_{\mu\nu},\end{equation}
where $R_{ab}$  denote the generators of the $R-$symmetry group $SO_{R}(6)$, while $R_{\mu\nu}$  are those of the euclidean {\it Lorentz} group $SO(4)$. Here, we have again split the ten-dimensional indices into two subsets: the greek ones range from $1$ to $4$ and the roman ones ($a,b,...$), which run from $1$ to $6$.

Exploiting the explicit form \eqref{pw} and \eqref{cito1} of the solutions for $\epsilon_{s_{i}}$ and  $\epsilon_{c_{i}}$, we find  the following expression for the reduced super-algebra
\be
\left\{\bar{Q},Q\right\}=4 R_0\ \ \ \ \ \ \ \ \ \ \ \left\{Q,Q\right\}=\left\{\bar{Q},\bar{Q}\right\}=0\ \ \ \ \left[R_0,Q\right]=\left[R_0,\bar{Q}\right]=0,
\ee
where 
\be
\bar{Q}=\frac{1}{\sqrt{2}}\left(Q_1+iQ_2\right)\ \ \ \ \ \ \ Q=\frac{1}{\sqrt{2}}\left(Q_1-iQ_2\right).
\ee
The bosonic generator $R_0$ is a linear combination of the $R-$symmetry and rotation generators. In terms 
of the couplings appearing in the Wilson loops  it is given by 
\be
R_0=\frac{1}{2}\Omega^{\mu\nu}\left(R_{\mu\nu}+M^a_{\,\, \mu}\left[M\left(M^TM\right)^{-1}\right]^b_{\,\, \nu}R_{ab}\right).
\ee
The reduced algebra is $SU(1|1)$. The entire super-algebra  is simply obtained by boosting up this one.

Next we consider the case when two columns of the matrix $M^{a}_{\ \mu}$ vanish and these Wilson loops are  $1/8$   BPS. Since only three {\it effective} scalars couple to  these second family of operators, there is  an obvious invariance under the   $SU(2)$   acting on the $R-$symmetry  directions orthogonal to these scalars. Then the four complex  supercharges can be organized  in two doublets $\{\tilde Q_{\alpha}\}$ ($\mathbf{2}$)  and 
$\{\tilde Q_{\bar\alpha}\}$ ($\mathbf{\bar 2}$)  of this $SU(2)$ and the reduced super-algebra  takes the form
\begin{equation}
\label{Alg}
\left\{\tilde{Q}_\alpha,\tilde{Q}_\beta\right\}=4\left(\overline{C \sigma^I}\right)_{\alpha\beta}R_I \ \ \ 
\ \ \ \  \left\{\tilde{Q}_{\bar{\alpha}},\tilde{Q}_{\bar{\beta}}\right\}=4\left(C \sigma^I\right)_{\bar{\alpha}\bar{\beta}}R_I\ \ \ \  \left\{\tilde{Q}_{\alpha},\tilde{Q}_{\bar{\beta}}\right\}=4\delta_{\alpha\bar{\beta}}R_0
\end{equation}
 where $C=i\sigma_2$ is the two-dimensional charge conjugation matrix.  The bosonic  part is instead given by
\be  
\left[R_0,R_I\right]=0\ \ \ \ \ \ \ \ \ \left[R_I,R_J\right]=i\epsilon_{IJK}R_K.
\ee
The action of the bosonic generators on the supercharges contains the obvious  transformation rule under the $SU(2)$ 
\begin{equation}
\label{Alg1}
\left[R_I,\tilde{Q}_\alpha\right]=-\frac{1}{2} \sigma^I_{\alpha\beta}\tilde{Q}_\beta \ \ \ \ \ \ \ \left[R_I,\tilde{Q}_{\bar{\alpha}}\right]=\frac{1}{2} \bar{\sigma}^I_{\bar{\alpha}\bar{\beta}}\tilde{Q}_{\bar{\beta}},
\ee
but also those under the $SO(2)$ generated  by $R_{0}$
\be
\left[R_0,\tilde{Q}_\alpha\right]=-\frac{1}{2} C_{\alpha\bar{\beta}}\tilde{Q}_{\bar{\beta}} \ \ \ \ \ \ \ 
\left[R_0,\tilde{Q}_{\bar{\alpha}}\right]=\frac{1}{2} C_{\bar{\alpha}\beta}\tilde{Q}_{\beta}.
\end{equation}
To provide an explicit form of the bosonic generators in terms of the Wilson loop couplings, we normalize the two non-vanishing columns of $M$ to obtain two orthonormal vectors 
\[
m_\mu^c={M^c}_\mu/\left({M^b}_\mu{M^b}_\mu\right)\ \ \ \ \ \mu=3,4.
\]
These vectors are orthogonal to $B$.  In order to complete the orthonormal basis, we have to add three complex vectors, which will denote with $n^{a}_{1}, n^{a}_{2}$ and $w^{a}$. Then  $SO\left(3\right)$ 
and $SO(2)$ symmetry associated to rotations of the scalar subspace spanned by $n_1,n_2,w$ are generated by  
\be
R_0=\frac{1}{2}\Omega^{\mu\nu}\left(R_{\mu\nu}+2\ \delta_{\mu3} \delta_{\nu4}  \,m_3^a\, m_4^b \,R_{ab}\right)
\ee
\be
R_1=-n_1^a\, n_2^b\,R_{ab}\ \ \ \ \ \ \ \ R_2=n_1^a\, w^b\, R_{ab}\ \ \ \ \ \ \ R_3=-n_2^a\, w^b\, R_{ab}.
\ee
This is the usual $SU(1|2)$, which  also appears in the case of  DGRT loops living on $S^{2}$.

\section{Perturbative aspects}
\label{Perb aspects}
In this Section we explore the quantum behavior at weak-coupling of the Lissajous Wilson loops: as we will see, the tight relation with the circular loop, that appears obvious at level of symmetries, will also become evident in the perturbative computation. 

We start by considering the familiar perturbative expansion of the Wilson loop, directly derived from its definition as path-ordered exponential (in the following we will consider the Wilson loop in the fundamental representation):
\begin{equation} <\mathcal{W}(\gamma)>=
\frac{1}{N}\sum_{n=0}^\infty
\oint_{\gamma} ds_1\int_0^{s_1}ds_2\cdots
\int_0^{s_{2n-1}}ds_{2n}  {\rm Tr}
\bigl<{\cal A}(s_1)
\cdots
{\cal A}(s_{2n})\bigr)\bigr>,
\end{equation}
where we have expressed the expansion in terms of correlators of the effective connection 
$$ {\cal A}(x(s))=A_\mu(x(s)) u^\mu(x(s))+\Phi_a (x(s))u^a(x(s)).$$In $\mathcal{N}=4$ SYM theory Wilson loops with smooth contours that are locally supersymmetric exhibit an improved ultraviolet behavior, 
making them manifestly finite in perturbation theory \cite{Er}. This property nicely shows up at the first non-trivial order of the perturbative expansion, being encoded in the particular structure of the effective propagator appearing in the computation (in Feynman gauge):

\begin{equation}\label{contr}
\langle {\cal A}^{A}(s_1){\cal A}^{B}(s_2)\rangle_0
=\frac{g^2}{4\pi^2}
\frac{u^a(s_1)u_a(s_2)+u^\mu(s_1)u_\mu(s_2)}{( x_{1}-x_{2})^2}\,\delta^{AB}.
\end{equation}
The finiteness of the first order contribution can be proved in full generality \cite{Bassetto:2008yf}, but additional surprising properties are manifest for globally supersymmetric loops. In order to proceed we have, as first step, to compute in our case the combined vector-scalar propagator, effectively attaching on the loop contour. Taking into account the explicit form of the scalar couplings $u^a$ in terms of $M$ and $B$ (see eq.(\ref{coupl1a})) and the relevant constraints (\ref{localsusy}), we obtain:
\begin{equation}
\frac{u^a(s_1)u_a(s_2)+u^\mu(s_1)u_\mu(s_2)}{( x_1-x_{2})^2}=\frac{B^2}{2}.
\label{propa}
\end{equation}
We recover therefore, for a general Lissajous loop, the very same result of the circular Wilson loops: in Feynman gauge, the relevant effective propagator appearing in the perturbative expansion, is constant when the initial and final points are attached on the loop. This peculiar property was taken originally as an indication that the path-integral computation of circular 1/2 BPS loops reduces to a matrix-model expectation value \cite{Er,Drukker:2000rr}, a fact that has been proved later by localization \cite{Pestun:2007rz}.  We find instructive to derive the above result also from a more general point of view, expressing the vector and the scalar couplings directly in terms of the Killing spinors associated to our loops.  
Let us consider, for a generic non-pure conformal Killing spinor $\epsilon\left(x\right)$, the structure of the bilinear $u^M(x)$ that defines the couplings: 
\[u^M\left(x\right)=\epsilon^{T} C^{-1}\gamma^{M}\epsilon=\epsilon_s\Gamma^M\epsilon_s+2x_\mu\epsilon_c\Gamma^{\mu M}\epsilon_s+2\delta^{M\mu}x_\mu\epsilon_c\epsilon_s+2x_\mu x_\nu\delta^{\nu M}\epsilon_c\Gamma^\mu\epsilon_c-x^2\epsilon_c\tilde{\Gamma}^M\epsilon_c.
\]
[In the last equalities, we have shifted  to the chiral notation introduced in App.\ref{app1} for Dirac matrices and spinors to have a more manageable notation.]
Because $u^M\left(x\right)u_M\left(x\right)=0$, we have that only quadratic terms in $x_1,x_2$ contribute, and the numerator of our effective propagator turns out to be:
 \begin{equation}\label{prop}u^M\left(x_1\right)u_M\left(x_2\right)=2\left(x_1-x_2\right)_\mu\left(x_1-x_2\right)_\nu \epsilon_c\Gamma^\mu\epsilon_c u^\nu\left(x_2\right)-\left(x_1-x_2\right)^2\ \epsilon_c\tilde{\Gamma}^M\epsilon_c u^M(x_2).\end{equation}
The second term in this expression can be rewritten using
$$\epsilon_c\tilde{\Gamma}^M\epsilon_c\ u_M\left(x_2\right)=\left(\epsilon_c\tilde{\Gamma}^M\epsilon_c\right)\left(\epsilon_s\Gamma_M\epsilon_s\right)+ 4\epsilon_c\epsilon_s \ x_{2\rho}\ \epsilon_c\Gamma^\rho\epsilon_c + 2x_{2\rho}x_{2\sigma} \epsilon_c\Gamma^\rho\epsilon_c\ \epsilon_c\Gamma^\sigma\epsilon_c,$$ 
and we see again that, in absence of the term related to special conformal transformations ($\epsilon_c\Gamma^\rho\epsilon_c=0$), the effective propagator results constant and coincides, of course, with (\ref{propa}). The fact that special conformal transformations could change the value of the propagator was already noticed in \cite{Drukker:2000rr} and it is at the very root of the difference between the expectation value of infinite lines and circular loops.

\begin{figure}[ht]
\centering
	\includegraphics[width=.5 \textwidth]{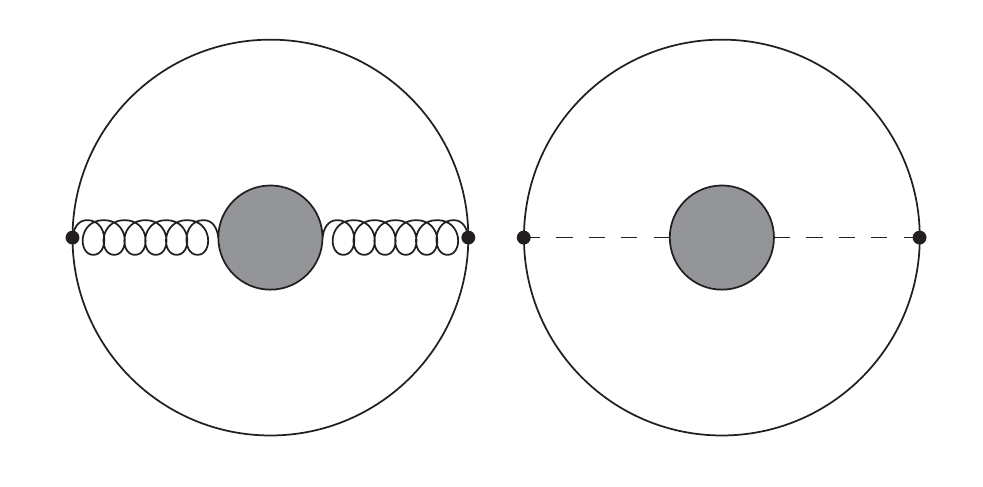}
	\put(-162,-5){(a)}	\put(-63,-5){(b)}
\caption{\label{bubble}One-loop correction to the gluon and the scalar exchange.}
\end{figure}

The first order contribution is 
\begin{equation}
\label{single-exch}
\mathcal{W}_1(\gamma)
= \frac{g^2 N}{8\pi^2} \oint_{\gamma} d s_1 ds_2
\frac{u^a(s_1)u_a(s_2)+u^\mu(s_1)u_\mu(s_2)}{(x_1-x_2)^2}\equiv\frac{g^2 N}{8\pi^2}\Sigma_2[\gamma].
\end{equation}
Because the periodicity of our loops is $2\pi n/\Omega_1$, as explained in Section 2, we obtain the same result of the circular loop up the replacement 
\begin{equation}
g^2\to -\frac{B^2 n^2}{\Omega_1^2} g^2, 
\label{rescaling}
\end{equation}
where we have taken into account the difference in the constant effective propagator. The next step is to compute the second non-trivial order in the perturbative expansion: at the order $g^4$, the different contributions are not separately finite and we have to introduce the regularization procedure. On the other hand, 
in the circular case, it was shown in \cite{Er} that, using dimensional regularization, divergencies cancel and the remaining finite pieces can be easily evaluated. The same behavior was recognized for generic DGRT loops on $S^2$ \cite{Bassetto:2008yf}, where a general and compact expression for the combined  one-loop corrected propagators and internal vertices was provided. Here we follow the same strategy: firstly,  we consider the effect of the one-loop correction to the effective propagator. The relevant diagrams are schematically displayed in fig. \ref{bubble}
and in the following we shall refer to them as  the {\it bubble diagrams}.

\begin{figure}[ht]
\centering
	\includegraphics[width=.5 \textwidth]{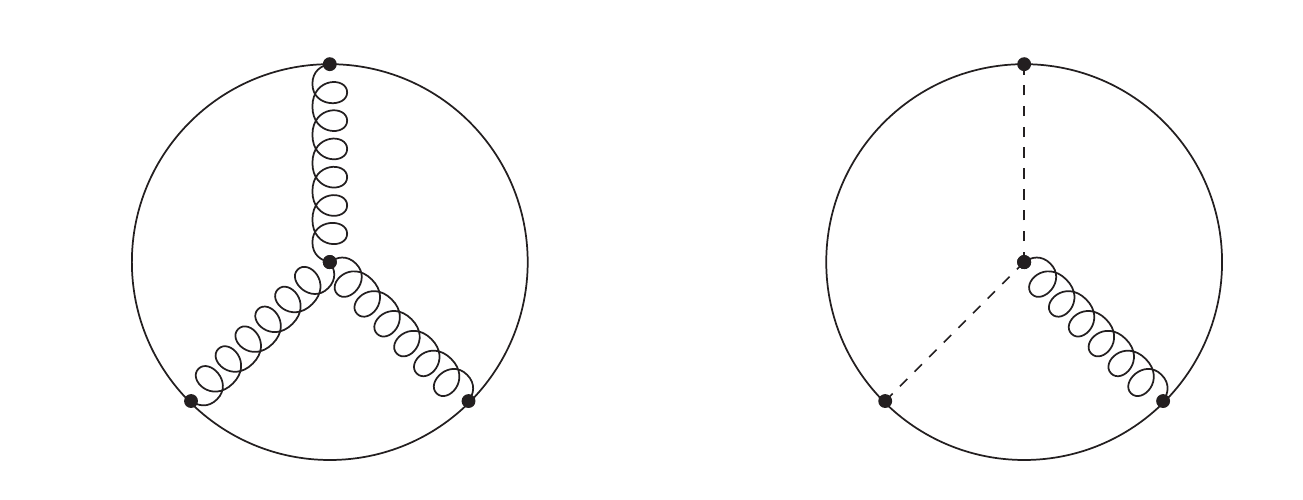}
	\put(-167,-5){(a)}	\put(-52,-5){(b)}
\caption{ \label{spider} Spider-diagrams: gauge and scalar  contribution}
\end{figure}

\noindent
In Feynman gauge they can be easily computed with the help of \cite{Er},
where the one-loop correction to the gauge and scalar propagator
has been calculated.  The final result is (here $D=2\omega$)
\begin{equation}
S_2=-g^4 (N^2-1)\frac{\Gamma^2(\omega-1)}
{2^{7}\pi^{2\omega}(2-\omega)(2\omega-3)} \Sigma_{4\omega-6}[\gamma],
\label{bubbleres}
\end{equation}
that clearly exhibits a pole at $\omega=2$. The next step, at this order, is to investigate the so-called \textit{spider diagrams}, namely the perturbative contributions coming from the gauge vertex $A^3$ and the scalar-gauge vertex $\phi^2 A$ (see fig. \ref{spider}).  We have to compute
\be
\label{spider1}
S_3=
\frac{g^3}{3 N}\oint_{\gamma} d{s}_1 d{s}_2 d{s}_3\eta({s}_1,{s}_2,{s}_3)
\langle \mathrm{Tr}[\mathcal{A}({s}_1)\mathcal{A}({s}_2)\mathcal{A}({s}_3)]\rangle_0,
\ee
 where
 \begin{equation}
\label{spider3}
\eta(s_1,s_2,s_3)=\theta(s_1-s_2)\theta(s_2-s_3)+\mathrm{cyclic\ permutations}.
\end{equation}
\noindent After a simple computation $S_3$ takes the form
\be
\label{spider4}
S_3=\frac{ g^4 (N^2-1)}{8} B^2\oint_\gamma ds_1 ds_2 ds_3~\epsilon(s_1,s_2,s_3) (x_1-x_3)^2
\dot{x}^\mu_2\frac{\partial \mathcal{I}_1(x_3-x_1,x_2-x_1)}{\partial x_3^\mu},
\ee
where we have introduced the symbol
\[
\epsilon(s_1,s_2,s_3)=\eta(s_1,s_2,s_3)-\eta(s_2,s_1,s_3),
\]
that is a totally antisymmetric object in the permutations of $(s_1,s_2,s_3)$ and its value is $1$ when $s_1>s_2>s_3$. The
quantity   $\mathcal{I}_1(x,y)$ is defined as the following integral in momentum space
\be
\label{I1}
\mathcal{I}_1(x,y)\equiv
\int \frac{d^{2\omega} p_1 d^{2\omega} p_2}{(2\pi)^{4\omega}}\frac{e^{i p_1 x+i p_2 y}}{p_1^2
p_2^2 (p_1+p_2)^2}.
\ee
Following closely the same steps in reference \cite{Bassetto:2008yf}, we can factor out from (\ref{spider4}) a contribution that completely cancels the divergent and finite part of the bubble, leaving us with a regular expression proportional to
$$S_2+S_3 \simeq \oint ds_1ds_2 ds_3 \ \epsilon(s_1,s_2,s_3)\frac{\left(x_3-x_2\right)\cdot\dot{x}_2}{\left(x_3- x_2\right)^2}	\log\left[{\frac{\left(x_2-x_1\right)^2}{\left(x_3-x_1\right)^2}}\right]$$
that in the parametrization (\ref{liss}) turns out to be
$$\oint_\gamma ds_1ds_2 ds_3 \ \epsilon\left(s_1,s_2,s_3\right)\frac{\Omega_1\cos^2\frac{\theta}{2} \sin\Omega_1(s_3 - s_2) +\Omega_2\sin^2\frac{\theta}{2}  \sin \Omega_2 \left(s_3 - s_2\right) }{2\left(1-\cos^2\frac{\theta}{2} \cos\Omega_1\left(s_3 - s_2)\right) -\sin^2\frac{\theta}{2}  \cos \Omega_2 \left(s_3 - s_2\right)\right)}\times$$	$$\times\log\left[{\frac{\left(1-\cos^2\frac{\theta}{2} \cos\Omega_1\left(s_2 - s_1\right) -\sin^2\frac{\theta}{2}  \cos \Omega_2 \left(s_2 - s_1\right)\right)}{\left(1-\cos^2\frac{\theta}{2} \cos\Omega_1\left(s_3 - s_1\right) -\sin^2\frac{\theta}{2}  \cos \Omega_2 \left(s_3 - s_1\right)\right)}}\right].$$

The integral is potentially complicated, but it actually vanishes because the integrand is antisymmetric in the exchange $s_2\leftrightarrow s_3$ while the measure and the integration domains are symmetric. This parallels exactly the circular case.

To get the complete two-loop answer we have still to consider the \textit{double-exchange} diagrams to the perturbative expansion of the Wilson loop, namely we have to analyze the contribution
\be
\frac{g^4}{N}\oint_\gamma d s_1 d s_2  d s_3  d s_4\theta(s_1-s_2)\theta(s_2-s_3)\theta(s_3-s_4)
\langle \mathrm{Tr}[\mathcal{A}(s_1) \mathcal{A}(s_2) \mathcal{A}(s_3) \mathcal{A}(s_4)]\rangle_0.
\ee
It is quite clear that, due to the constant character of the effective propagator, we simply recover again the circular result up the rescaling (\ref{rescaling}): we are led therefore to conjecture that the exact quantum expectation value of the Lissajous Wilson loop is simply obtained from the circular Wilson loop, once  (\ref{rescaling}) is taken into account

\begin{equation}
<\mathcal{W}(\gamma)>
=\frac{1}{N}L_{N-1}^1\left(g^2\,\frac{B^2 n^2}{\Omega_1^2}\right)
\exp\left[-\frac{g^2}{2}\,\frac{B^2 n^2}{\Omega_1^2}\right]\,.
\label{lissresult}
\end{equation}

The original derivation of the above formula \cite{Er,Drukker:2000rr} relied mainly on the assumption that interacting contributions vanish to all order in perturbation theory: the resummation of all the exchange diagrams can be easily performed, a job effectively done by a Gaussian matrix-model. The ultimate reason of the correctness of such derivation stands, of course, on the proof, given by Pestun \cite{Pestun:2007rz}, that the full ${\cal N}=4$ path-integral computing the 1/2 BPS circular loop reduces to the matrix-model, exploiting a localization procedure. In our case we have not been able to prove a similar result and we can only provide further evidences for the conjecture. Taking the large $N$ limit and defining $\lambda=g^2 N$ we get

\begin{equation}\label{expv}\langle \mathcal{W}\rangle=\frac{2}{\sqrt{\lambda\delta}}I_1\left(\sqrt{\lambda\delta}\right)\end{equation} where 
$\delta=-B^2 n^2/\Omega_1^2$ and $I_1$ is the Bessel function. Due to AdS/CFT correspondence, we expect that as $\lambda$ becomes large this result, if correct, should match a classical string theory solution, $i.e$ we should recover on the string side the behavior

\begin{equation}\label{expvncl}\langle \mathcal{W}\rangle\sim\exp\left(\sqrt{\lambda\delta}\right).\end{equation}  

We will see indeed the matching in next section.

\section{Strong coupling: classical solution}
In the following we shall  discuss the string theory duals  of this  family of  loops.  There is an obvious issue that  we have to address before proceeding: the scalar couplings $u^{a}$ are in general complex, hence the usual interpretation of $u^{a}$ as a  6-component vector drawing a contour in $S^{5}$ is apparently lost.  A similar situation was considered in  \cite{arXiv:0902.4586},  where the strong coupling regime of loops lying on a hyperbolic sub-manifold  of the space-time was investigated. There,  it was suggested that 
the dual open string does not  move in the usual $S^{5}$ sphere, but in its complexification. In other words,  if  $Y^{a}$, with $Y^{a} Y^{a}=1$,  are the six flat  coordinates  spanning the sphere, they must be allowed to assume complex values: $y^{i}\in \mathds{C}^{6}$. 

This prescription is not uncommon in the AdS/CFT correspondence: a typical situation arises when considering charged local operators in the Euclidean theory. In this case, studying for example operators like $\mathrm{Tr} [Z^{J}]$ (the BMN ground state \cite{BMN}), we have   a semiclassical description of these objects, in the Lorentzian theory, in terms of particle trajectories or giant gravitons. Of course in the Euclidean theory there is no time and real propagations from the boundary of $AdS_5$ into the bulk are not available. It was suggested  (see the discussion in \cite{Yoneya:2003mu}) that considering a tunneling picture or, equivalently, a complexification of the space, the problem can be resolved and it basically corresponds to Wick-rotate one of the directions on $S^{5}$.

We come now to examine our specific problem: determining  the classical string solution dual to the  $1/16$ BPS Lissajous  Wilson loops.  To construct it, we have to {\it minimize} the Polyakov action
\begin{equation}
\label{sigma model0}
S=\frac{\sqrt{\lambda}}{4\pi}\int d^2\sigma\sqrt{h}h^{\alpha\beta} G_{MN}\left(X\right)\partial_\alpha X^N\partial_\beta X^M,
\end{equation}
with the boundary conditions fixed by the Wilson loop.  Here $G_{MN}$ is the $AdS_5\times S^5$ metric, while  $h^{\alpha\beta}$ is the world-sheet metric. In what follows we shall use the  conformal gauge and we shall set $\sqrt{h}h^{\alpha\beta}=\delta^{\alpha\beta}$.  Since we are dealing with a single loop operator, 
 the $AdS_5$ and the $S^5$ parts of the sigma model are  completely decoupled and they can be solved separately. In particular the Virasoro constraints of the two sectors must be satisfied independently. 
 
To begin with, we shall discuss  the euclidean $AdS_{5}$  sector of the  $\sigma-$model,   following closely \cite{Drukker:2007qr,Drukker:2005cu}  where a general techniques for investigate toroidal loops is presented.  Firstly, they parametrize the $AdS_5$ metric as follows 
\be
ds^2=-dr_0^2+r_0^2dv^2+dr_1^2+r_1^2d\phi_1^2+dr_2^2+r_2^2d\phi_2^2
\ee
where the radial coordinates $r_{i}$ obey the constraint
\be
-r_0^2+r_1^2+r_2^2=-1,
\ee
to ensure that we are describing euclidean $AdS_{5}$. Since the boundary conditions are set by a Lissajous figure of the form\footnote{We have redefined  the solution for mere convenience.}
\be
\gamma\left(\tau\right): x^{\mu}(\tau)=\left\{\cos \frac{\theta}{2}\cos \Omega_{1}\tau,\ \cos \frac{\theta}{2}\sin \Omega_{1}\tau,\ \sin \frac{\theta}{2}\sin \Omega_2\tau,\ \sin \frac{\theta}{2}\cos \Omega_2\tau\right\},
\ee
a natural ansatz for describing the world-sheet is provided by
\be
r_i=r_i\left(\sigma\right),\ \ \ \ v=v_{0} \ \ ({\rm const.)},\ \ \ \ \phi_1=\Omega_2\tau\ {\rm and}\ \ \phi_{2}=\Omega_{1}\tau.
\ee
Then the reduced action for the remaining dynamical variable  can be written as
\begin{equation}
\label{actionAds}
S=\frac{\sqrt{\lambda}}{4\pi}\int d\sigma \left(-r_0'^2+r_1'^2+r_2'^2+\Omega_2^2r_1^2+\Omega_1^2r_2^2+\Lambda\left(-r_0^2+r_1^2+r_2^2+1\right)\right)
\end{equation}
where $\Lambda$ is a Lagrange multiplier and the prime denotes the derivative with respect to $\sigma$. Since we are working in the conformal gauge, this action must be  supplemented with  the Virasoro constraint
\be
-r_0'^2+r_1'^2+r_2'^2-\Omega_2^2r_1^2-\Omega_1^2r_2^2=0.
\ee
In \cite{Drukker:2007qr,Drukker:2005cu}  it was pointed out that the dynamics described by the action \eqref{actionAds} is integrable  and hence one can find  a complete set of  integrals of motion.  For example 
one can consider
\be
\begin{split}
I_{0}=& r_0^2-\frac{1}{\Omega_1^2}\left(r_0r_1'-r_1r_0'\right)^2-\frac{1}{\Omega_2^2}\left(r_0r_2'-r_2r_0'\right)^2=1,\\
I_{1}=&r_1^2-\frac{1}{\Omega_1^2}\left(r_0r_1'-r_1r_0'\right)^2+\frac{1}{\left(\Omega_1^2-\Omega_2^2\right)}\left(r_1r_2'-r_2r_1'\right)^2=0,
\end{split}
\ee
together with the Virasoro constraint.  We can now  change variables from $r_{i} \ (i=1,2,3)$  to $\left(\zeta_1,\zeta_2\right)$ 
 \begin{equation}
 \label{chVar}
 r_0=\frac{\zeta_1\zeta_2}{\Omega_1\Omega_2}\ \ \ \ \ \ r_1=\sqrt{\frac{\left(\zeta_1^2-\Omega_2^2\right)\left(\zeta_2^2-\Omega_2^2\right)}{\Omega_2^2\left(\Omega_1^2-\Omega_2^2\right)}}\ \ \ \ \ \ \ r_2=\sqrt{\frac{\left(\zeta_1^2-\Omega_1^2\right)\left(\zeta_2^2-\Omega_1^2\right)}{\Omega_1^2\left(\Omega_2^2-\Omega_1^2\right)}}
 \end{equation} 
with $\Omega_2\leqslant\zeta_1\leqslant \Omega_1\leqslant\zeta_2$.  The equation of motion for these new unknowns  can be then determined from the integral of motions and they are given by
\be
\zeta_{1}^{\prime}=\pm \frac{\zeta_{1}^{2}-\Omega_{2}^{2}}{\zeta_{1}^{2}-\Omega_{1}^{2}} \ \ \ \ \ \ 
\zeta_{2}^{\prime}=\pm \frac{\zeta_{2}^{2}-\Omega_{2}^{2}}{\zeta_{2}^{2}-\Omega_{1}^{2}}.
\ee
Actually we do not need to solve explicitly the equations of motion to determine the value of the classical 
action on the solutions. Exploiting again the integral of motions,  the action turns out to be
\be
\label{minAds}
\begin{split}
S_{AdS_{5}}=&\frac{\sqrt{\lambda}}{2\pi}\int d\sigma d\tau\left(\Omega_2^2r_1^2+\Omega_1^2r_2^2\right)=-\frac{\sqrt{\lambda}}{2\pi}\int d\sigma d\tau\left(\zeta_1'+\zeta_2'\right)=\\
=&-\frac{\sqrt{\lambda}}{2\pi}\left(\int_{\frac{\Omega_2\Omega_1}{\sqrt{\Omega_2^2\sin^2\frac{\theta}{2}+\Omega_1^2\cos^2\frac{\theta}{2}}}}^{\Omega_2} d\zeta_1+\int_\infty^{\Omega_1} d\zeta_2\right)\int d\tau=\\
\simeq&-\frac{\sqrt{\lambda}}{2\pi}\left(\Omega_2+\Omega_1-{\frac{\Omega_2\Omega_1}{\sqrt{\Omega_2^2\sin^2\frac{\theta}{2}+\Omega_1^2\cos^2\frac{\theta}{2}}}}\right)\int d\tau,
\end{split}
\end{equation} 
where in the last expression the divergence was removed by hand. The integration domains are determined by (\ref{chVar}); for $r_1$ and $r_2$ the integration is from the boundary to the interior of $AdS$: at the boundary we have $r_1,r_2\rightarrow\infty$ with $\frac{r_1}{r_2}\rightarrow \tan\frac{\theta}{2}$, while in the interior $r_1,r_2\rightarrow0$.  The integration over $\tau$ is trivial and it simply produces the range of this variable: $\tau=\left(0,2\pi n/\Omega_{1}\right)$.  Summarizing we have obtained
\begin{equation}\label{minAdsL}S_{AdS_{5}}=-\frac{n}{\Omega_1}\sqrt{\lambda}\left(\Omega_1+\Omega_2-{\frac{\Omega_1\Omega_2}{\sqrt{\Omega_1^2\cos^2\frac{\theta}{2}+\Omega_2^2\sin^2\frac{\theta}{2}}}}\right).\end{equation}

Next we consider the $S^{5}$ sector of the $\sigma-$model. We again prefer to use euclidean flat coordinates and to write the action as follows
\begin{equation}
\label{sigma model}
S_{S^5}=\frac{\sqrt{\lambda}}{4\pi}\int d^2\sigma [ \partial_\alpha Y^{a}\partial^{\alpha} Y^{a}-\Lambda
(Y^{a} Y^{a}-1)],
\end{equation}
where $a$ run from $1$ to $6$ and $\Lambda$ is a Lagrange multiplier which ensures that the target space is a {\it sphere}.  In order to simplify the explicit form of the boundary conditions to be imposed, we introduce a new set of coordinates defined by 
\be
y_{0}=b^{a} Y_{a},\ \ \  y_{5}= b^{a}_{0} Y_{a}\ \ \ \mathrm{and}\ \ \   y_{\mu}= m^{a}_{\,\mu} Y^{a},
\ee 
where $b^{a}$ is the normalized vector $B^{a}/\sqrt{(B\cdot B)}$, $m^{a}_{\, \mu}$  are the four orthonormal vectors given by $m^{a}_{\,\mu}=M^{a}_{\,\mu}/\sqrt{(M_{\mu}\cdot M_{\mu})}$
and finally $b^{\,a}_{0}$ is a sixth vector of unit norm and orthogonal to the previous ones. At the level
of the gauge theory, this corresponds to a (complex) redefinition of the scalars such that the matrix $M^{a}_{\, \mu}$ of couplings is diagonal.

Since  the above transformation is an element of $SO(6,\mathds{C})$, the action \eqref{sigma model} is 
substantially unchanged.  The original  normalized vector $u^{a}$ of scalar couplings  takes
 the following simpler form  
\be
\begin{split}
\Theta_0= b^{a} u_{a}= &\frac{\sqrt{-B\cdot B}}{\sqrt{\Omega_1^2\cos^2 \frac{\theta}{2}+\Omega_2^2\sin^2 \frac{\theta}{2}}},\ \ \ \ \ \ \ \ \ \ \ \ 
\Theta_5= b_{0}^{\,a}u_{a}=0,\\
\Theta_{4}+i \Theta_{3}=&m^{a}_{\ 4}u_{a}+i m^{a}_{\ 3}u_{a} =- i\frac{\sin \frac{\theta}{2}\sqrt{-B\cdot B-\Omega_2^2}~e^{i\Omega_2\tau}}{\sqrt{\Omega_1^2\cos^2 \frac{\theta}{2}+\Omega_2^2\sin^2 \frac{\theta}{2}}},\\
\Theta_{1}+i \Theta_{2}=&m^{a}_{\ 1}u_{a}+i m^{a}_{\ 2}u_{a} =-i\frac{ \cos \frac{\theta}{2}\sqrt{-B\cdot B-\Omega_1^2}~e^{i\Omega_1\tau}}{\sqrt{\Omega_1^2\cos^2 \frac{\theta}{2}+\Omega_2^2\sin^2 \frac{\theta}{2}}},
\end{split}
\ee
in this new basis.
Next we perform a further change of variables and instead of the four $y_{\mu}$  we introduce the toroidal coordinates
\be
r_{1} e^{i\phi_{1}}=y_{4}+i y_{3}\ \ \ \ \mathrm{and}\ \ \ \  r_{2} e^{i\phi_{2}}=y_{1}+i y_{2}.
\ee
In term of these coordinates, the $\sigma-$model action can be cast into the form
\begin{equation}
\label{sigma model1}
\begin{split}
S_{S^5}=\frac{\sqrt{\lambda}}{4\pi}\int d^2\sigma&\biggl[ \partial_\alpha y_{0}\partial^{\alpha} y_{0}+\partial_\alpha y_{1}\partial^{\alpha} y_{1}+\partial_\alpha r_{1}\partial^{\alpha} r_{1} +r_{1}^{2}
\partial_\alpha \phi_{1}\partial^{\alpha} \phi_{1} +\partial_\alpha r_{2}\partial^{\alpha} r_{2} +\\
&\ \ \ \ \ + r_{2}^{2}
\partial_\alpha \phi_{2}\partial^{\alpha} \phi_{2} -\Lambda
((y^{0})^{2} +(y^{5})^{2} +r_{1}^{2}+r_{2}^{2}-1)\biggr].
\end{split}
\end{equation}
The boundary conditions for the $S^{5}$ sector are then simply given by\footnote{We choose $\sqrt{-B\cdot B}$ and  $\sqrt{-B\cdot B-\Omega_{1,2}^2}$ to be real. This choice will simplify our treatment. }
\be
\begin{split}
y_0=&\frac{\sqrt{-B\cdot B}}{\sqrt{\Omega_1^2\cos^2 \frac{\theta}{2}+\Omega_2^2\sin^2 \frac{\theta}{2}}},\ \ \ \ \ \ \ \ \ \ \ \ 
y_5=0,\\
r_1 e^{i\phi_1}=&-i\frac{\sin \frac{\theta}{2}\sqrt{-B\cdot B-\Omega_2^2}~e^{i\Omega_2\tau}}{\sqrt{\Omega_1^2\cos^2 \frac{\theta}{2}+\Omega_2^2\sin^2 \frac{\theta}{2}}},\ 
r_2 e^{i\phi_2}=-i\frac{ \cos \frac{\theta}{2}\sqrt{-B\cdot B-\Omega_1^2}~e^{i\Omega_1\tau}}{\sqrt{\Omega_1^2\cos^2 \frac{\theta}{2}+\Omega_2^2\sin^2 \frac{\theta}{2}}}.
\end{split}
\ee
The above equality are meant to hold when $\sigma$ reaches its boundary value. Since we prefer to have boundary conditions, which are (formally) real, we perform another Wick-rotation: $y_{k}\mapsto i y_{k}$
for $k=1,\dots, 5$. After these procedure $S^{5}$ becomes euclidean $AdS_{5}$ and we can choose the 
following ansatz  for the  solutions
\be
y_{0}=0,\ \ \  y_5=0\ \ \ \ \ \ \ r_i=r_i\left(\sigma\right)\ \ i=1,2\ \ \ \ \ \ \ \phi_1=\Omega_2\tau\ \ \ \ \phi_2=\Omega_1\tau.
\ee
With this choice, apart from  an overall sign in the action and for the boundary conditions,  we are left with the same problem encountered in the $AdS_5$  sector.  Thus  we can follow the same procedure and introduce the new coordinates $\left(\hat{\zeta}_1,\hat{\zeta}_2\right)$
\be
y_0=\frac{\hat{\zeta}_1\hat{\zeta}_2}{\Omega_1\Omega_2}\ \ \ \ \ \ r_1=\sqrt{\frac{\left(\hat{\zeta}_1^2-\Omega_2^2\right)\left(\hat{\zeta}_2^2-\Omega_2^2\right)}{\Omega_2^2\left(\Omega_1^2-\Omega_2^2\right)}}\ \ \ \ \ \ \ r_2=\sqrt{\frac{\left(\Omega_1^2-\hat{\zeta}_1^2\right)\left(\hat{\zeta}_2^2-\Omega_1^2\right)}{\Omega_1^2\left(\Omega_1^2-\Omega_2^2\right)}}
\ee
with $\Omega_2\leqslant\hat{\zeta}_1\leqslant \Omega_1\leqslant\hat{\zeta}_2$. The action has again the simple form
\be
S=\frac{n}{\Omega_1}\sqrt{\lambda}\int d\sigma \left(\hat{\zeta}_1'+\hat{\zeta}_2'\right).
\ee
The integration is from the boundary values to $r_1,r_2=0$. For the latter we have $\hat{\zeta}_1=\Omega_2$ and $\hat{\zeta}_2=\Omega_1$.  The boundary values at infinity are trickier to obtain, but after some algebraic manipulations,  the boundary condition on $r_{i}$ translates into
\be
\hat{\zeta}^2_1=-B \cdot B\ \ \ \ \ \ \ \hat{\zeta}^2_1=\frac{\Omega_1^2\Omega_2^2}{\Omega_1^2\cos^2 \frac{\theta}{2}+\Omega_2^2\sin^2 \frac{\theta}{2}}
\ee
or
\be
\hat{\zeta}^2_2=\frac{\Omega_1^2\Omega_2^2}{\Omega_1^2\cos^2 \frac{\theta}{2}+\Omega_2^2\sin^2 \frac{\theta}{2}}\ \ \ \ \ \ \ \hat{\zeta}^2_2=-B \cdot B.
\ee
The choice between the two possibilities depends on the values of the parameters $B,\Omega_1,\Omega_2,\theta$, but this does not affect the result for the action:
\be
S_{S^{5}}=\frac{n}{\Omega_1} \sqrt{\lambda} \left(\Omega_1+\Omega_2-\sqrt{-B\cdot B}-\frac{\Omega_1\Omega_2} { \sqrt{\Omega_1^2 \cos^2 \frac{\theta}{2}+\Omega_2^2\sin^2 \frac{\theta}{2} } }\right).
\ee
The total result is obtained by adding the two contributions: $S_{AdS_{5}}+S_{S^{5}}$. We have
\be
S=-\frac{n}{\Omega_1} \sqrt{\left(-B\cdot B\right)\lambda} .
\ee
As anticipated in the previous Section, we have been able to reproduce by AdS/CFT correspondence the strong-coupling result conjectured from the perturbative computation: we notice that the expression suggested from the weak coupling expansion originates from a precise cancellation between the $AdS_5$ and the $S^5$ sectors of the $\sigma$-model. Another remark concerns the subtractions we have done by hands to get a finite action: the very same result would be obtained by applying the Legendre transformation procedure, proposed in \cite{Drukker:2000rr}, that is generally considered the correct one. We feel therefore quite confident that an exact localization underlies our results.

\section*{Acknowledgements}
This work was supported in part by the MIUR-PRIN contract 2009-KHZKRX. We are pleased to thank Marco Bertolini and Fabrizio Pucci for participating at the early stage of the project. We also thank Antonio Bassetto and Nadav Drukker for useful discussions.

\newpage
\appendix
 \begin{flushleft}
 {\huge \bf Appendices}\hfill
 \end{flushleft}
\addcontentsline{toc}{section}{\large Appendices}
\renewcommand{\theequation}{\Alph{section}.\arabic{equation}}
\section{General framework and Conventions}
\label{app1}
{\sc Dirac Algebra and spinors in D=10:} The Euclidean Dirac algebra in ten dimensions is defined by the anti-commutation rules
\be
\gamma^{M}\gamma^{N}+\gamma^{N}\gamma^{M}=2 \delta^{MN}\mathds{1},
\ee
where the $\gamma^{M}$ are $32\times 32$ matrices. We shall use the Weyl representation, where
the chiral operator,
$
\gamma^{11}=-i\gamma^{1}\gamma^{2}\cdots\gamma^{10},
$
is diagonal and it is given by 
\be
\gamma^{11}=\begin{pmatrix} \mathds{1}_{16\times 16} & 0\\ 0 &  -\mathds{1}_{16\times 16}\end{pmatrix}.
\ee
With this choice we can write the Dirac matrices $\gamma^{M}$ in block form
\be
\gamma^{M}=\begin{pmatrix} 0 & \tilde \Gamma^{M}\\ \Gamma^{M} & 0\end{pmatrix}.
\ee
The non-vanishing blocks are related by hermitian conjugation
$
\tilde\Gamma^{M}=(\Gamma^{M})^{\dagger}
$
and they obey the  chiral algebra
\be
\Gamma^{M}\tilde\Gamma^{N}+\Gamma^{N}\tilde\Gamma^{M}=2 \eta^{MN}\mathds{1}\ \ \ \ 
\mathrm{and}\ \ \ \ \ 
 \tilde\Gamma^{M}\Gamma^{N}+\tilde\Gamma^{N}\Gamma^{M}=2 \eta^{MN}\mathds{1}.
\ee
In addition they can be taken symmetric, {\it i.e.} $(\Gamma^{M})^{t}=\Gamma^{M}$ and $(\tilde\Gamma^{M})^{t}=\tilde\Gamma^{M}$.  Differently from what occurs in the case of Lorentz signature
we cannot choose the  above blocks to be also real, we can only impose
\be
\label{real}
\left\{\begin{matrix} (\Gamma^{M})=(\Gamma^{M})^{*}\\
(\tilde\Gamma^{M})=(\tilde\Gamma^{M})^{*}
\end{matrix}\right.  \ \ \ \  \mathrm{for}\ \ \ M\ne 10 \ \ \ \mathrm{and}\ \ \ \ \ 
\left\{\begin{matrix}  (\Gamma^{10})^{*}=-\Gamma^{10}\\
 (\tilde\Gamma^{10})^{*}=-\tilde\Gamma^{10}.
 \end{matrix}\right.
\ee
The reality condition \eqref{real} combined with the previous requirements  will also imply that $\tilde \Gamma^{M}=\Gamma^{M}$ for $M\ne 10$ and $\tilde\Gamma^{10}=-\Gamma^{10}$.

An explicit representation for these blocks can be constructed by identifying the $\Gamma^{M}$ with the real representation of the Euclidean Clifford algebra in $9$ dimensions  for $M\ne 10$  and by posing $\Gamma^{10}=i\mathds{1}$.

Finally the matrices $\Gamma^{M}$ and $\tilde \Gamma^{M}$  obey two important Fierz identities which play a key role in any computation involving  ten dimensional supersymmetry
\be
\label{Fierz}
(\Gamma^{M})_{\alpha_{1}(\alpha_{2}}(\Gamma_{M})_{\alpha_{3}\alpha_{4})}=0\ \ \ \ \ 
\mathrm{and}\ \ \ \ \  (\tilde\Gamma^{M})_{\alpha_{1}(\alpha_{2}}(\tilde\Gamma_{M})_{\alpha_{3}\alpha_{4})}=0
\ee
where $\alpha_{1}, \alpha_{2},\alpha_{3},\alpha_{4} = 1,\dots,16$  are the matrix indices of $\Gamma^{M}$ and $\tilde\Gamma^{M}$.

In this representation the $32-$component Dirac spinor $\psi$ naturally splits into  two  Weyl spinors of opposite chirality with 16 components each
\be
\psi=\begin{pmatrix} \psi_{+}\\ \psi_{-}\end{pmatrix}.
\ee
The blocks $\Gamma^{M}$ and $\tilde\Gamma^{M}$ will act on $\psi_{+}$ ($\Gamma^{M}\psi_{+}$) and $\psi_{-}$ ($\tilde\Gamma^{M}\psi_{-}$)  respectively by reversing their chirality. Since $\Gamma^{M}$ and $\tilde\Gamma^{M}$ coincide for $M\ne 10$, we shall not distinguish between them when it is not necessary
and for instance we shall sometimes write $\Gamma^{\mu}\psi_{-}$  (with $\mu=1,\dots,4$)  instead of
$\tilde\Gamma^{\mu}\psi_{-}$.

It is also convenient to introduce  the matrices
\be
\Gamma^{MN}=\frac{1}{2}(\tilde \Gamma^{M}\Gamma^{N}-\tilde \Gamma^{N}\Gamma^{M})\ \ \ \ 
\mathrm{and}\ \ \ \ \tilde \Gamma^{MN}=\frac{1}{2}(\Gamma^{M}\tilde \Gamma^{N}-\Gamma^{N}\tilde \Gamma^{M}),
\ee
which are proportional to the generators of the two irreducible chiral representations of the rotation group
and preserve chirality.  

In euclidean space there is no need  for complex conjugation to write down a fermionic invariant bilinear. In the 32-component notation we can in fact  introduce the complex contraction
$\psi C^{-1}\psi$\ \mbox{where} \mbox{$C=-i\gamma^{10}$ is the charge conjugation matrix}  $[ C^{-1}\gamma^{M}C=-(\gamma^{M})^{t}]$.

In the Weyl representation  this combination  can be  equivalently written in terms of the 16-component Weyl spinors as
follows
\be
\widetilde\psi\psi=2 \psi_{+}\psi_{-}.
\ee
This  unusual choice for fermionic  bilinears is important when considering supersymmetric theories
in euclidean space since it avoids the introduction of $\psi^{\dagger}$ in the construction 
of an invariant action. 

{ \sc Euclidean ${\cal N}$=4 SYM:}
Let us discuss the euclidean version of $\mathcal{N}=4$ SYM on a flat $d=4$ space-time, with gauge group $G$.  As we have extensively done in this paper, the theory can be viewed  as  the  dimensional reduction of $\mathcal{N}=1$ SYM in $d=10$ and  its action takes the following compact form,
\begin{equation}
\label{action}
S=-\frac{1}{2g^2_{YM}}\int d^4xTr\left(\frac{1}{2}F_{MN}F^{MN}-\Psi \Gamma^MD_M\Psi\right)\ \ \ \ \ \ \ N,M=1...10,
 \end{equation}
when  using  the ten-dimensional notation \footnote{In the following we shall closely follow the convention of \cite{Dymarsky:2009si}.}. The matrices $\Gamma^M$ have been introduced in the previous subsection. In eq. \eqref{action} all fields take value in the Lie algebra of $G$ while
 the covariant derivative and the field strength are
given by
$$D_M\equiv\partial_M+A_M\ \ \ \ \ \ \ \ F_{MN}\equiv\left[D_M,D_N\right].$$
Denoting  the space-time directions with greek indices $\mu,\nu=1...4$  and  the remaining ones with $A,B=5...10$, 
the bosonic part of the familiar ${\cal N}=4$ Lagrangian   emerges by expanding  \eqref{action} in terms of $\left(A_\mu,\Phi_A\right)\equiv A_M.$


\noindent
The superconformal transformations, which leave invariant the action of $\mathcal{N}=4$ SYM, in ten dimensional  notation are given by
\begin{equation}
\label{traconf }
\delta A_M=\epsilon\left(x\right)\Gamma_M\Psi\ \ \ \ \ \ \delta\Psi=\frac{1}{2}F_{MN}\Gamma^{MN}\epsilon\left(x\right)-2\Phi_A\tilde{\Gamma}^A\epsilon_c.
\end{equation}
The parameter $\epsilon\left(x\right)$ is a conformal Killing spinor in flat space, {\it i.e.}
\begin{equation}
\label{Kspinor}
\epsilon\left(x\right)=\epsilon_s+x_\mu\Gamma^\mu\epsilon_c,\end{equation} where $\epsilon_s$ and $\epsilon_c$ are two constant spinors. In particular $\epsilon_s$ has positive (ten-dimensional) chirality and it generates the usual  16 Poincar\'e supersymmetries, while  $\epsilon_c$ is anti-chiral and yields the remaining 16 fermionic superconformal symmetries.  

In the subsequent appendices  shall use  the above chiral notation $(\Gamma^{M},\tilde{\Gamma}^{M})$, while in the main text we have  adopted for  simplicity the standard $32\times 32$ Dirac matrices $\gamma^{M}$. 

\section{The classification of the impure Wilson loops}
\label{Aapp}
In this appendix we give a general classification of the possible impure Wilson loops, constructing explicitly both the contours and the scalar couplings. In doing so, we heavily relies on the properties of the Killing spinors associated to the loops and on the well-known six-dimensional embedding of the conformal group. Our strategy is to reduce the classification to independent representatives up to conformal transformations and to solve properly the basic constraints in the different cases. The general situation is reached by "conformally boosting" the relevant contours and couplings, using the appropriate conformal transformations described in appendix \ref{app3}.
\subsection{General Properties}
\label{GP}
We start the construction of an {\it impure} Wilson loop $\mathcal{W}(\gamma)$ with assigning a conformal Killing spinor,
\be
\epsilon=\epsilon_{s}+x^{\mu}\Gamma_{\mu}\epsilon_{c},
\ee
for which the vector  $\epsilon\Gamma^{M}\epsilon\ne 0$
 does not vanish and in particular its space-time components, $\epsilon\Gamma^{\mu}\epsilon$, define a real four-vector with unit norm\footnote{We can weaken this condition and require that $\epsilon\Gamma^{\mu}\epsilon$  is proportional to a real vector, since $\epsilon$ is defined up to  the scaling $\epsilon\mapsto \lambda \epsilon$ with $\lambda\in \mathds{C}^{*}$  }. Then the contour $\gamma$  is  obtained by solving the differential equation
\be
\label{pino}
\begin{split}
\dot{x}^{\mu}=\epsilon\Gamma^{\mu}\epsilon=&~\epsilon_{s}\Gamma^{\mu}\epsilon_{s}+2\epsilon_{s}\epsilon_{c} x^{\mu}+2 
\epsilon_{s}\tilde\Gamma^{\mu}_{\ \nu} \epsilon_{c}x^{\nu}+2x^{\mu}  x^{\nu}\epsilon_{c}\tilde\Gamma_{\nu}\epsilon_{c}-x^{2}\epsilon_{c}\tilde\Gamma^{\mu}\epsilon_{c}=\\
=&~a^{\mu}+\lambda x^{\mu}+\Omega^{\mu}_{\  \ \nu} x^{\nu}+2 x^{\mu} (b\cdot x) -x^{2} b^{\mu}.
\end{split}
\ee
On the r.h.s of \eqref{pino} we recognize an infinitesimal generator  of the conformal algebra, which is 
a combination of  a translation $(a^{\mu}\equiv 
\epsilon_{s}\Gamma^{\mu}\epsilon_{s})$,  a dilation $(\lambda\equiv 2
\epsilon_{s}\epsilon_{c})$, a rotation $(\Omega^{\mu}_{\ \ \nu}\equiv 
2\epsilon_{s}\tilde\Gamma^{\mu}_{\  \  \nu}\epsilon_{c})$  and a special conformal transformation
$(b^{\mu}\equiv \epsilon_{c}\tilde\Gamma^{\mu}\epsilon_{c})$.  In other words, as we have anticipated in the main text,  $x^{\mu}(s)$ is
an orbit of the conformal group.

\noindent
Its explicit form  can be easily constructed in terms of the solutions of  the six-dimensional linear system
\be
\label{Csect2}
\dot{Y}^{m}(s)=W^{m}_{\ \ n} Y^{n}(s) \ \ \ \   (m,n=1,\dots,6) ,
\ee
where constant matrix $W$  realizes the canonical embedding of the (euclidean) conformal transformation  in the r.h.s of \eqref{pino} into the algebra of $SO(5,1)$, {\it i.e.}
\be
\label{WWW}
W\equiv
\begin{pmatrix}
0 & \Omega^{1}_{\ \ 2 }& \Omega^{1}_{\ \ 3 } & \Omega^{1}_{\ \ 4 } &  a^{1}+b^{1}
&  a^{1}-b^{1}\\
- \Omega^{1}_{\ \ 2 }& 0  & \Omega^{2}_{\ \ 3 } & \Omega^{2}_{\ \ 4 } &  a^{2}+b^{2}
&  a^{2}-b^{2}\\
- \Omega^{1}_{\ \ 3 } & -\Omega^{2}_{\ \ 3 } & 0&\Omega^{3}_{\ \ 4 } &  a^{3}+b^{3}
&  a^{3}-b^{3}\\
- \Omega^{1}_{\ \ 4 } & -\Omega^{2}_{\ \ 4} &-\Omega^{3}_{\ \ 4 } & 0 & a^{4}+b^{4}
&  a^{4}-b^{4}\\
-a^{1}-b^{1} & -a^{2}-b^{2} & -a^{3}-b^{3} & -a^{4}-b^{4} & 0  & -\lambda\\ 
a^{1}-b^{1} & a^{2}-b^{2} & a^{3}-b^{3} &a^{4}-b^{4} & -\lambda & 0 
\end{pmatrix}.
\ee
In fact we can write that 
\be
\label{curve1}
x^{\mu}(s)=\frac{Y^{\mu}(s)}{Y^{5}(s)+Y^{6}(s)}\ \ \ \  (\mu=1,\dots,4),
\ee
where  $Y^{m}(s)$  is  subject to  the initial condition  $Y^{m}(0)=y^{m}_{0}$ with $(y_{0}\cdot y_{0})=0$.

The scalar couplings of  the {\it impure} Wilson loops  are instead parametrized, in general,  by two constant six-dimensional vectors $B^{a}_{0}$ and $B^{a}_{1}$ and a $6\times 4$ rectangular matrix $M^{a}_{\  \ \mu}$, the vector $u^{a}=\epsilon\Gamma^{a}\epsilon$ (with $a=1,\dots,6$) in \eqref{WLL} being naturally arranged as follows
\be
\label{couplA}
\begin{split}
u^{a}=&\epsilon\Gamma^{a}\epsilon=(\epsilon_{s}\Gamma^{a}\epsilon_{s}-x^{2}
\epsilon_{c}\tilde\Gamma^{a}\epsilon_{c})+2\epsilon_{s}\Gamma^{a}\tilde\Gamma_{\nu}\epsilon_{c} x^{\nu}\equiv\\
\equiv & B^{a}_{0}-x^{2} B^{a}_{1}+M^{a}_{\  \ \mu} x^{\mu}.
\end{split}
\ee
Remarkably, the coefficients  $B^{a}_{0},\ B^{a}_{1}$ and  $M^{a}_{\  \ \mu}$ in \eqref{couplA} are not free quantities but they are subject to some constraints which keep track of their {\it spinorial} origin.  For example the Fierz identity
\be
2(\epsilon_{s}\Gamma_{M}\epsilon_{s})(\epsilon_{s}\Gamma^{M}\tilde\Gamma_{\nu}\epsilon_{c})=0
\ee
translates into
\be
\label{cons1}
a_{\nu}\Omega^{\nu}_{\ \ \mu}+\lambda a_{\mu}+ B^{a}_{0} M^{a}_{\ \ \mu}=0
\ee
and similarly we can also show 
\begin{subequations}
\label{cons2}
\begin{align}
\label{cons2a}
& b_{\nu}\Omega^{\nu}_{\ \ \mu}+B^{a}_{1} M^{a}_{\ \ \mu}-\lambda b_{\mu}=0,\\
\label{cons2b}
& (a\cdot a)+(B_{0}\cdot B_{0})=(b\cdot b)+(B_{1}\cdot B_{1})=0,\\
\label{cons2c}
& M^{a}_{\ \ \mu} M^{a}_{\ \ \nu}+(\lambda \delta^{\rho}_{\mu}+\Omega^{\rho}_{\ \ \mu})(\lambda\delta^{\rho}_{\nu}+
\Omega^{\rho}_{\ \ \nu})+2
(a_{\mu} b_{\nu}+a_{\nu} b_{\mu})=2\delta_{\mu\nu}[ (a\cdot b)+(B_{0}\cdot B_{1})].
\end{align}
\end{subequations}
The converse is also true: {\it i.e.}  if the couplings $M^{a}_{\ \ \mu}$, $B^{a}_{0}$ and $B^{a}_{1}$
obey the constraints \eqref{cons1} and \eqref{cons2}, they define a supersymmetric  Wilson loop.

In Subsec. \ref{vuop0} we shall discuss in detail  the possible solutions of the above set of constraints.

\subsection{Classification of the possible orbits}
\label{vuop}
The  circuits for the {\it impure} Wilson loops are provided  by orbits of the conformal group: we shall write down all the relevant contours modulo conformal equivalence, {\it i.e.} we shall identify loops which differ by a conformal transformation. It is convenient to rephrase this problem in six dimensional language and to  list all the possible forms of the matrix \eqref{WWW} up to the  adjoint action of an element of  $SO(5,1)$.  This classification is achieved by separating the matrices  $W$ into  two classes  according to the value of their determinant.

\paragraph{\sc  ${\bf det(W)\ne 0:}$} 
\noindent

In this case the kernel of $W$ is trivial and the matrix possesses 6 eigenvalues  different from zero which can be generically paired in three couples:
\be
(i\Omega_{1}, -i\Omega_{1}),\ \ (i\Omega_{2}, -i\Omega_{2})\ \ \  \mathrm{and}\ \ \  (\rho, -\rho),
\ee
with $\Omega_{1,2}$ and $\rho$ real numbers.  The eigenvectors associated to $\rho$ and $-\rho$ are two real independent light-like vectors with respect to $SO(5,1)$ invariant metric $\eta=\mathrm{diag}(1,1,1,1,1-1)$.  The linear space orthogonal to these eigenvectors is an invariant subspace 
and on such subspace $W$ is an anti-hermitian matrix defining the generator of an $SO(4)$ rotation.

Thus   the matrix \eqref{WWW} can be always cast in the following canonical form 
\be
\label{WWWC}
W_{C}\equiv\begin{pmatrix} 0 & \Omega_{1} & 0 & 0 & 0& 0\\
- \Omega_{1} & 0 & 0 & 0 & 0& 0\\
0 & 0 & 0 & \Omega_{2} & 0& 0\\
0 & 0 & - \Omega_{2} & 0 & 0& 0\\
0 & 0 & 0 & 0 & 0& \rho\\
0 & 0 & 0& 0 & \rho& 0\\
\end{pmatrix},
\ee
up to an $SO(5,1)$ transformation. The corresponding contour is then given by
\be
\label{spiral}
\begin{array}{ll}
x^{1}= e^{\rho s} \cos\theta_{0} \cos(\Omega_{1}s)\ \  \ \ \  \ \  \ \ \ & x^{3}= e^{\rho s}\sin\theta_{0} \cos(\Omega_{2} s)\\
x^{2}=  e^{\rho s}\cos\theta_{0}  \sin(\Omega_{1} s)\ \ \ \ \ \ \ \ \ \ \  & x^{4}= e^{\rho s}\sin\theta_{0}  \sin(\Omega_{2} s)\\
\end{array}
\ee
For $\rho>0$ it is an infinite open path that  starts from the origin ($x^{\mu}=0$) at $s=-\infty$ and it reaches  infinity when $s=\infty$. It is obtained by composing two planar motions on orthogonal planes. The motion
on each  plane  is the well-known {\it logarithmic spiral.}

\paragraph{\sc  ${\bf det(W)=0:}$}  
\noindent

In this case the kernel of $W$ is not empty and we have three different  possibilities.

\medskip
{\sc {\bf (A)} } The kernel of $W$ includes a {\it time-like vector}: $W$ defines a rotation and up to an  $SO(5,1)$ transformation 
the matrix  \eqref{WWW} takes the form
\be
\label{WWWCA}
W_{C}\equiv\begin{pmatrix} 0 & \Omega_{1} & 0 & 0 & 0& 0\\
- \Omega_{1} & 0 & 0 & 0 & 0& 0\\
0 & 0 & 0 & \Omega_{2} & 0& 0\\
0 & 0 & - \Omega_{2} & 0 & 0& 0\\
0 & 0 & 0 & 0 & 0& 0\\
0 & 0 & 0& 0 & 0& 0\\
\end{pmatrix}.
\ee
The contour  is simply given by
\be
\begin{array}{ll}
x^{1}=  \cos\theta_{0} \cos(\Omega_{1}s)\ \  \ \ \  \ \  \ \ \ & x^{3}= \sin\theta_{0} \cos(\Omega_{2} s)\\
x^{2}=  \cos\theta_{0}  \sin(\Omega_{1} s)\ \ \ \ \ \ \ \ \ \ \  & x^{4}= \sin\theta_{0}  \sin(\Omega_{2} s),
\end{array}
\ee
and it lies on a sphere $S^{3}$. It is generically an open orbit but it closes if the ratio $\Omega_{1}/\Omega_{2}\in \mathds{Q}$ ({\it Lissajous figures}). This class of contours are discussed in details in the main text.  

{\sc {\bf (B)} } The kernel of $W$ contains a {\it light-like vector} $v_{\ell}$: up to an $SO(5,1)$ we can always choose $v_{\ell}=(0,0,0,0,-1,1)$ and the matrix $W$ can be rearranged as follows
\be
\label{WWWCB1}
W_{C}\equiv\begin{pmatrix} 0 & \Omega_{1} & 0 & 0 & a_{1}& a_{1}\\
- \Omega_{1} & 0 & 0 & 0 & a_{2}& a_{2}\\
0 & 0 & 0 & \Omega_{2} & a_{3}& a_{3}\\
0 & 0 & - \Omega_{2} & 0 & a_{4}& a_{4}\\
-a_{1} & -a_{2} & -a_{3} &-a_{4} & 0& 0\\
a_{1} & a_{2} & a_{3}& a_{4} & 0& 0\\
\end{pmatrix},
\ee
where we have used the residual $SO(4)$ invariance of $v_{\ell}$ to put $\Omega^{\mu}_{\ \ \nu}$ in
its canonical form. If $\Omega_{1}$ and $\Omega_{2}$ are different from zero, we can further set $a_{i}=0$ by
means of the adjoint action of a translation, {\it i.e.} we have again obtained the canonical form \eqref{WWWCA}.

If only one  of the $\Omega_{i}$ vanishes, {\it e.g.} $\Omega_{2}=0$, we can always reduce $W$ to
\be
\label{WWWCB2}
W_{C}\equiv\begin{pmatrix} 0 & \Omega_{1} & 0 & 0 & 0& 0\\
- \Omega_{1} & 0 & 0 & 0 & 0&0\\
0 & 0 & 0 &0 & 0& 0\\
0 & 0 & 0 & 0 & a& a\\
0 & 0 & 0 &-a & 0& 0\\
0 & 0 & 0& a & 0& 0\\
\end{pmatrix},
\ee
 by means of a translation in the plane  $(1,2)$ and of a rotation in the plane $(3,4)$. Thus the corresponding 
 contour is an {\it helix} whose parametric equation is
\be
\begin{array}{ll}
x^{1}=  \cos\theta_{0} \cos(\Omega_{1}s)\ \  \ \ \  \ \  \ \ \ & x^{3}= x^{3}_{0}\\
x^{2}=  \cos\theta_{0}  \sin(\Omega_{1} s)\ \ \ \ \ \ \ \ \ \ \  & x^{4}=a t.
\end{array}
\ee
If both  the $\Omega_{i}$ vanish,  $W$ generates a translation and its canonical form is given by 
\eqref{WWWCB2} for $\Omega_{1}=0$. The path is then given by a straight-line parallel to 
the  {\it coordinate axis} $x^{4}$.

{\sc {\bf (C)} } The kernel consists only of {\it space-like vectors}: we can easily show that its dimensions is at least 2
and thus  $W$ reduces to a generator of $SO(n,1)$ (with $n=1,3$) on the sub-space orthogonal to the kernel.

For $n=1$ we have a pure dilation, namely  we obtain $\eqref{WWWC}$ with $\Omega_{1}=\Omega_{2}=0$. The contour is again a {\it straight-line} given by 
\be
x^{\mu}=x^{\mu}_{0} e^{\lambda t}.
\ee
 Instead for $n=3$ we have a dilation and a planar rotation, {\it i.e.} \eqref{WWWC}  with $\Omega_{1}=0$. The contour is obtained from \eqref{spiral} by posing $\Omega_{1}=0$.
 
  \subsection{Classification of the couplings}
\label{vuop0}
We investigate here the structure of the scalar couplings for the orbits classified in Subsec.\ref{vuop}.  More precisely, we shall show how to solve the constraints \eqref{cons1} and \eqref{cons2} and  to parameterize the different solutions. 

\paragraph{\sc ${\bf det(W)=0}$:} 
\noindent

${\bf (A)}$ We consider first the case when $W_{C}$ is a pure rotation. 
Since $a^{\mu}$ and $b^{\mu}$ vanish,  $B_{0}^{a}$ and 
$B_{1}^{a}$ are  $6-$component  light-like complex vectors because of  \eqref{cons2b}. Let us assume that 
$(B_{0}\cdot B_{1})\ne 0$, then, for instance, $B_{0}^{a}$ can be put in the following
canonical form\footnote{A non-vanishing light-like vector with respect to the euclidean metric is complex.
Its real and imaginary part defines two real orthogonal vectors of equal norm. The form \eqref{B31} is a trivial consequence of these properties.}
\be
\label{B31}
B_{0}^{a}=p_{1}(0,0,0,0, 1 , i),
\ee
up to an $SO(6)$ rotation. The light-like vector $B^{a}_{1}$ can be instead parameterized as follows
\be
\label{B1}
B_{1}^{a}=\left(\hat B_{1}^{1},\hat B_{1}^{2},\hat B_{1}^{3},\hat B_{1}^{4}, \frac{1}{2}
\left[m- \frac{\hat B_{1}^{i}\hat B_{1}^{i}}{m}\right],\frac{i}{2}
\left[m+ \frac{\hat B_{1}^{i}\hat B_{1}^{i}}{m}\right]\right).
\ee
The  special form of $B_{0}$ reduces the condition \eqref{cons1} to 
\be
M^{5}_{\,\, \mu}+ i M^{6}_{\,\, \mu} =0,
\ee
which is solved by  setting
\be
M^{5}_{\,\, \mu}=p_{1}q_{\mu}\ \ \ \ \mathrm{and}\ \ \  \ \ M^{6}_{\,\,  \mu}= i  p_{1}q_{\mu},
\ee
where $q$ is a  four-vector fixed by the condition \eqref{cons2a}. We get
\be
\label{B24}
q_{\mu}=-\frac{1}{(B_{0}\cdot B_{1})}\sum_{i=1}^{4}\hat B_{1}^{i}\hat M^{i}_{\,\, \mu},
\ee
where $\hat M^{i}_{\,\, \mu}$ is $4\times 4 $ matrix obtained from  $M^{a}_{\,\, \mu}$ by erasing 
the last two rows.  We remark that $\frac{p_{1}}{m} (\hat B_{1}^{i}\hat B_{1}^{i})=-(B_{0}^{a} B_{1}^{a})$.
Finally, we have to consider the quadratic condition \eqref{cons2c}
\be
\sum_{a=1}^{4}\hat M^{i}_{\,\, \mu} \hat M^{i}_{\,\, \nu}+\Omega^{\rho}_{\,\, \mu}
\Omega^{\rho}_{\,\, \nu}
=2\delta_{\mu\nu}(B_0\cdot B_1),
\ee
which in turn implies
\be
\label{vv}
\sum_{a=1}^{4}\hat M^{i}_{\,\, \mu} \hat M^i_{\ \ \nu}=\mbox{\small $\begin{pmatrix}
2 (B_0\cdot B_1)-\Omega_{1}^{2} & 0 & 0 &0\\
0& 2 (B_0\cdot B_1)-\Omega_{1}^{2}  & 0 &0\\
0& 0 & 2 (B_0\cdot B_1)-\Omega_{2}^{2}  &0\\
0 & 0 & 0 &2 (B_0\cdot B_1)-\Omega_{2}^{2}
\end{pmatrix}$}.
\ee
The general solution of \eqref{vv} is provided by
\be 
\label{ww}
\hat M^{i}_{\,\, \mu}=S\begin{pmatrix} a_{1} & 0 & 0 &0\\
0 & a_{1} & 0 &0\\
0 & 0 &  a_{2}  &0\\
0  & 0 &  0 & a_{2}
\end{pmatrix},
 \ee
where $S$ is a matrix of $SO(4,\mathds{C})$ and $a_{i}^{2}={2 (B_0\cdot B_1)-\Omega_{i}^{2}}$.
Equivalently we can say that  the columns of $\hat{M}^{a}_{\,\, \mu}$ define four complex orthogonal vectors whose norms
are given by the diagonal element in the r.h.s. of  eq. \eqref{vv}.  

An apparent singular point  in our analysis occurs when either $\Omega_{1}^{2}$ or $\Omega_{2}^{2} $ are equal to $2 (B_{0}\cdot B_{1})$. Let us consider, for instance, the case $\Omega_{1}^{2}=2 (B_{0}\cdot B_{1})$: the first two columns of the reduced matrix 
$\hat M^{i}_{\,\, \mu}$ are two light-like orthogonal complex vectors and up to an $SO(4)$ rotation we can set
\be
\hat M^{i}_{\,\, 1}=(im_{1},m_{1},0,0).
\ee
In turn the vector $\hat M^{i}_{\,\, 2}$ 
can be taken of the form
\be
\hat M^{i}_{\,\, 1}=(i m_{2}, m_{2}, i n_{2}, n_{2}),
\ee
and the remaining  two columns can be parameterized as follows
\be
\begin{split}
\hat M^{i}_{\,\, 3}=&(i m_{3}, m_{3}, \sqrt{\Omega_{1}^{2}-\Omega_{2}^{2}}\cos\alpha, \sqrt{\Omega_{1}^{2}-\Omega_{2}^{2}}\sin\alpha)\\
\hat  M^{i}_{\,\, 4}=&(i m_{4}, m_{4}, \sqrt{\Omega_{1}^{2}-\Omega_{2}^{2}}\cos\beta, \sqrt{\Omega_{1}^{2}-\Omega_{2}^{2}}\sin\beta).
 \end{split}
\ee
If $\Omega^{2}_{1}\ne 
\Omega^{2}_{2}$,
the orthogonality between $\hat M^{i}_{\,\, 2}$ and $\hat M^{i}_{\,\, 3,4}$ implies $n_{2}=0$, while $(\hat M_{ 3}\cdot \hat M_{4})=0$ is equivalent to  $\beta-\alpha=\frac{\pi}{2}$. We end up with  the following matrix 
\be
\hat M_{\,\, \mu}^{i}=\begin{pmatrix} i m_{1} &  i m_{2} &  i m_{3} & i m_{4}\\
 m_{1} &  m_{2} &   m_{3}&  m_{4}\\ 0  & 0 & \sqrt{\Omega_{1}^{2}-\Omega_{2}^{2}}\cos\alpha &
-  \sqrt{\Omega_{1}^{2}-\Omega_{2}^{2}}\sin\alpha\\
0 & 0 & \sqrt{\Omega_{1}^{2}-\Omega_{2}^{2}}\sin\alpha &  \sqrt{\Omega_{1}^{2}-\Omega_{2}^{2}}\cos\alpha.
\end{pmatrix},
\ee
up to an $SO(4)$ rotation, and the four-component vectors  $\hat M^{i}_{\,\, 1}$ and $\hat M^{i}_{\,\, 2}$ 
are not only light-like but also  parallel. The same property is actually shared by the complete first two columns of 
$M^{a}_{\,\, \mu}$: by means of  \eqref{B24}, we can easily check that $M^{a}_{\,\, 1}=m_{1}  V^{a}$
and  $ M^{a}_{\,\, 2}=m_{2} V^{a} $, where
\be
\label{B32}
\begin{split}
V^{a}=\left (i,1,0,0, -\frac{p_{1}}{ (B_{0}\cdot B_{1}) }(i \hat B_{1}^{1}+\hat B_{1}^{2}) ,
-\frac{i p_{1} }{ (B_{0}\cdot B_{1}) }(i \hat B_{1}^{1}+\hat B_{1}^{2}) \right) .
\end{split}
\ee
The remaining two columns of the matrix $M^{a}_{\,\, \mu}$ can be instead reorganized  as follows
\begin{subequations}
\label{SS}
\begin{align}
\label{SSa}
M^{a}_{\,\, 3}=&m_{3} V^{a}+\mbox{\small$ \sqrt{\Omega_{1}^{2}-\Omega_{2}^{2}}\left(0,0,\cos\alpha,\sin\alpha,-\frac{p_{1}(\hat B_{1}^{3}\cos\alpha+\hat B_{1}^{4}\sin\alpha)}{(B_{0}\cdot B_{1})}, -\frac{i p_{1}(\hat B_{1}^{3}\cos\alpha+\hat B_{1}^{4}\sin\alpha)}{(B_{0}\cdot B_{1})}\right)$}=\nonumber
\\
=& m_{3} V^{a}+S^{a}_{\,\,  3},\\
\label{SSb}
M^{a}_{\,\, 4}=&m_{4} V^{a}+ \mbox{\small $ \sqrt{\Omega_{1}^{2}-\Omega_{2}^{2}}\left(0,0,-\sin\alpha,\cos\alpha,-\frac{p_{1}(\hat B_{1}^{4}\cos\alpha-\hat B_{1}^{3}\sin\alpha)}{(B_{0}\cdot B_{1})},-\frac{i p_{1}(\hat B_{1}^{4}\cos\alpha-\hat B_{1}^{3}\sin\alpha)}{(B_{0}\cdot B_{1})}\right)$}=\nonumber\\
=& m_{4} V^{a}+S^{a}_{\,\,  4},
\end{align}
\end{subequations}
where $S_{3}$ and $S_{4}$ are orthogonal to $V$.

If $\Omega_{1}^{2}=\Omega_{2}^{2}=2 (B_{0}\cdot B_{1})$ we find, instead, the following matrix
\be
\hat M_{\,\, \mu}^{i}=\mbox{\small $\begin{pmatrix} i m_{1} &  i m_{2} &  i m_{3} & i m_{4}\\
 m_{1} &  m_{2} &   m_{3}&  m_{4}\\ 0  & i n_{2} & i n_{3}&
i n_{4}\\
0 & n_{2} & n_{3} & n_{4}
\end{pmatrix}$}.
\ee

When $(B_{0}\cdot B_{1})=0$,  the above analysis  is not substantially altered if either $B_{0}^{a}$ or 
$B^{a}_{1}$ do not vanish.  For example, if we consider again the case $B_{0}^{a}\ne 0$ (with $\Omega_{1}\ne0$ and $\Omega_{2}\ne 0$),   the four vectors 
$M^{i}_{\,\, \mu}$ are still orthogonal and the condition  \eqref{cons2a} implies
\be
\sum_{i=1}^{4}\hat B_{1}^{i}\hat M^{i}_{\ \ \mu}=0\ \ \ \ \  \Rightarrow\ \ \ \ \ \hat B_{1}^{i}=0:
\ee
in other words,  the vector $B_{1}^{a}$ is parallel to $B_{0}^{a}$. 
In this case,  the vector $q_{\mu}$ is undetermined.

If  $B_{0}^{a}=B_{1}^{a}=0$  the only surviving constraint is 
\be
M^{a}_{\,\, \nu} M^{a}_{\,\, \nu}=-\begin{pmatrix}
\Omega_{1}^{2} & 0 & 0 &0\\
0& \Omega_{1}^{2}  & 0 &0\\
0& 0 & \Omega_{2}^{2}  &0\\
0 & 0 & 0 &\Omega_{2}^{2}
\end{pmatrix},
\ee
which simply states that the columns of $M^{a}_{\ \ \nu}$ are  six-component complex orthogonal vectors. This exhausts all the possibilities contained in the case ${\bf (A)}$.

${\bf (B)}$ We have to analyze the canonical form \eqref{WWWCB2} for the matrix $W$.  The vector $B_{1}^{r}$ is still light-like  and complex and, if different from the null vector, it can be chosen to be
 \be
\label{B311}
B_{1}^{a}=p_{1}(0,0,0,0, 1 , i),
\ee
up to an $SO(6)$ rotation. The  condition \eqref{cons2a} again implies 
\be
M^{5}_{\,\, \mu}+ i M^{6}_{\,\, \mu} =0,
\ee
which is solved by  setting
\be
\label{uu}
M^{5}_{\,\, \mu}=p_{1}q_{\mu}\ \ \ \ \mathrm{and}\ \ \  \ \ M^{6}_{\,\,  \mu}= i  p_{1}q_{\mu}.
\ee
The vector $q_{\mu}$ is now determined by the orthogonality conditions with respect to $B_{0}$ and one obtains
\be
\label{uu1}
 q_{\mu}=-\frac{1}{(B_{0}\cdot B_{1})}\sum_{i=1}^{4}\hat B_{0}^{i}\hat M^{i}_{\,\, \mu}.
\ee
The quadratic condition \eqref{cons2c} is substantially unaltered with respect to the case ${\bf (A)}$  and in fact  it can be arranged in the following 
way
\be
\label{vv1}
\sum_{a=1}^{4}\hat M^{i}_{\ \ \mu} \hat M^i_{\ \ \nu}=\mbox{\small $\begin{pmatrix}
2 (B_0\cdot B_1)-\Omega_{1}^{2} & 0 & 0 &0\\
0& 2 (B_0\cdot B_1)-\Omega_{1}^{2}  & 0 &0\\
0& 0 & 2 (B_0\cdot B_1)  &0\\
0 & 0 & 0 &2 (B_0\cdot B_1)
\end{pmatrix}$}.
\ee
The discussion of  its solution is very similar to the previous case.

\noindent 
If $B_{1}^{a}$ is identically zero, the vector $B_{0}^{a}$ cannot vanish since it is time-like and
it  can be chosen to be
\be
\label{B0}
B_{0}^{a}=\left(0,0,0,0, \frac{1}{2}
\left[m- \frac{a^{2}}{m}\right],\frac{i}{2}
\left[m+ \frac{a^{2}}{m}\right]\right)
\ee
up to an $SO(6)$ rotation. The  orthogonality condition  \eqref{cons1}  now implies 
\be
M^{5}_{\,\, \mu}=\frac{1}{2}
\left(m+ \frac{a^{2}}{m}\right)q_{\mu}\ \ \ \ \mathrm{and}\ \ \  \ \ M^{6}_{\,\,  \mu}= \frac{i}{2}
\left(m- \frac{a^{2}}{m}\right)q_{\mu}.
\ee
The quadratic condition now  takes the form
\be
\sum_{a=1}^{4}\hat M^{i}_{\,\, \mu} \hat M^{i}_{\,\, \nu}+\Omega^{\rho}_{\,\, \mu}
\Omega^{\rho}_{\,\, \nu}+a^{2} q_{\mu} q_{\nu}
=0,
\ee
whose general solution is provided by
\be
\hat M^{j}_{\,\, \nu}=i S^{j}_{\ \ r}\left(\Omega^{r}_{\,\, \nu}+ \frac{a}{P^{2}} P^{r} q_{\nu}\right),
\ee
where $P^{r}$ is a vector of the kernel of $\Omega$ and $S$ is a matrix of $SO(4,\mathds{C})$.

\paragraph{\sc ${\bf det(W) \ne 0}$ :} 
\noindent

This case is not really different from the case ${\bf (A)}$ for $\det(W) =0$. We have only a redefinition of the $a_{i}$'s in \eqref{ww}: $$a_{i}^{2}=2 (B_0\cdot B_1)-\Omega_{i}^{2}-\lambda^{2}. $$

\subsection{Construction of the Killing spinors generating the Wilson loops}
\label{esa}
We shall focus our attention here on the general case  $\epsilon_{c}\tilde\Gamma^{M}\epsilon_{c}\ne 0$\footnote{Equivalently we can say that the vector $B_{1}^{a}$ does not vanish.}. The other possibilities can be discussed in a similar way.

Since  the four vector $b^{\mu}$ vanishes for all the "fundamental" orbits considered in the subsect. \ref{vuop},  the ten dimensional vector
$\epsilon_{c}\tilde\Gamma^{M}\epsilon_{c}$ (with   $M=1,2,\dots,9,10$)  can be always put  in the canonical form 
\be
\label{epsc}
 \epsilon_{c}\tilde\Gamma^{M}\epsilon_{c}=(0,0,0,0,0,0,0,0,\underset{9}{1},-\underset{10}{i}),
\ee
modulo an $R-$symmetry rotation and a dilation. Then the Fierz identities \eqref{Fierz} allow us to translate eq. \eqref{epsc} into an equivalent and simpler statement for the spinor $\epsilon_{c}$, namely $\epsilon_{c}$ is
an eigen-spinor  of positive chirality  of the matrix $\tilde\Gamma^{9}$. For future convenience,
we shall decompose $\epsilon_{c}$ in eigenstates of $\hat{\Gamma}^5\equiv\tilde\Gamma^1\cdots\tilde\Gamma^4$ and we shall write
\be
\label{es}
\epsilon_c=\cos t \ \epsilon_c^++\sin t \ \epsilon_c^- \ \ \ {\rm where}  \ \ \ \ \epsilon_c^+\epsilon_c^+=\epsilon_c^-\epsilon_c^-=1.
\ee

We then proceed to construct the  spinor $\epsilon_{s}$.

 \paragraph{\sc ${\bf det(W)\ne 0}$ :} 
\noindent

By imposing  that $W$ has the canonical form \eqref{WWWC}, we  find that $\epsilon_{s}$ admits the following expansion
\be
\label{epsilons}
\begin{split}
\epsilon_{s}=&\frac{1}{2}\left(\mathds{1}+\sum_{i=1}^{4}\frac{\hat B_{0}^{i}~\Gamma_{i+4}}{ (B_{0}\cdot B_{1})} \right)
\left[\sec(t)\left(\lambda_{+} \mathds{1}+\Omega_{-}\tilde\Gamma^{12}\right)\epsilon_{c}^{+}+
\csc(t)\left(\lambda_{-} \mathds{1}+\Omega_{+}\tilde\Gamma^{12}\right)\epsilon_{c}^{-}\right]
,
\end{split}
\ee
where $\lambda_{\pm}=\frac{\rho\pm\sigma}{2}$, $\Omega_{\pm}=\frac{\Omega_{1}\pm\Omega_{2}}{2}$, $\sigma$ is an arbitrary real number and $\hat B^{i}_{0}$ are four complex arbitrary entries. 

We can now evaluate the remaining parameters in the scalar couplings.
 We  first compute  the constant vector $B_{0}$ and we obtain 
\be
\label{B00}
B_{0}\equiv \left(\hat B_0^1,\hat B_0^2,\hat B_0^3,\hat B_0^4,\frac{1}{2} \left((B_{1}\cdot B_{0})-\frac{|B_{0}|^2}{(B_{1}\cdot B_{0})}\right),\frac{i}{2} 
   \left((B_{1}\cdot B_{0})+\frac{|B_0|^2}{(B_{1}\cdot B_{0})}\right)\right)
\ee
with $|B_0|^2=(\hat B_{0}^{1})^{2}+(\hat B_{0}^{2})^{2}+(\hat B_{0}^{3})^{2}+(\hat B_{0}^{4})^{2}$. In eq. 
\eqref{B00} the symbol $(B_{1}\cdot B_{0})$ is actually a short-hand notation for the following combination of the parameters
\be\frac{1}{2} \left(\csc ^2(t) \left(\lambda_-^2+\Omega_{+}^2\right)+\sec ^2(t) \left(\lambda_+^2+\Omega_-^2\right)\right),
\ee 
but it also denotes its meaning in terms of the Wilson-loop couplings.
Next we shall determine the matrix $M^{a}_{\,\, \mu}$. The first four rows ( $i=1,\dots,4$) can be summarized  in the following expression
\be
\label{MAmu}
\begin{split}
\hat M^{i}_{\,\, \mu}=&(\lambda_{+}\tan(t)-\lambda_{-}\cot(t))\epsilon_{c}^{+}\Gamma^{i+4}\Gamma_{\mu}\epsilon_{c}^{-}-\Omega_{+}\cot(t)
\epsilon_{c}^{+}\Gamma^{i+4}\Gamma_{\mu}\Gamma^{12}\epsilon_{c}^{-}-\\
&-\Omega_{-}\tan(t)
\epsilon_{c}^{-}\Gamma^{i+4}\Gamma_{\mu}\Gamma^{12}\epsilon_{c}^{+}.
\end{split}
\ee
The  expression for  the  remaining two rows is not particularly elegant, but at the end we simply find that 
they are given by  eqs. \eqref{uu} and \eqref{uu1}. 

In this framework the   $SO(4,\mathds{C})$ freedom in constructing the matrix  $\hat M^{a}_{\,\, \mu}$, emphasized in the previous subsection,  is translated into the freedom to choose the spinor $\epsilon_{c}$
without altering the other couplings.  This arbitrariness obviously corresponds to the complex rotation in
the directions $(5,6,7,8)$.

\paragraph{\sc ${\bf det(W)= 0}$ :} 
\noindent

${\bf (A)}$ This case is obtained from the previous analysis
by setting $\rho=0$.   
\noindent

${\bf (B)}$ Also this case requires small changes. Apart from setting
to zero $\rho$ and $\Omega_{2}$ in \eqref{epsilons} we have to add a term proportional to $a$:
\be
a\frac{ \sin (2 t) }{\sigma ^2+\Omega_1^2}
\left[\sin(t)\left(
\sigma\Gamma^{4}\epsilon^{+}_{c}+ \Omega_{1}\Gamma^{3}\epsilon^{+}_{c}\right) -\cos(t)\left(\sigma\Gamma^{4}\epsilon^{-}_{c}+\Omega_{1}\Gamma^{3}\epsilon^{-}_{c}\right)\right].
\ee

On the contrary the matrix $\hat M^{i}_{\,\,  \mu}$ is unaffected by the new parameter $a$ and it is 
  simply obtained from \eqref{MAmu} by  posing $\Omega_{2}$ and $\rho$ to zero.  
  
\noindent
${\bf (C)}$ It does not requires new ingredient with respect to the
cases ${\bf (A)}$ and ${\bf (B)}$ and it can be obtained from them by choosing  some of the free parameters
to be zero.

 \section{Conformal transformations}
 \label{app3}
In the previous appendix we have briefly discussed the possible {\it impure} Wilson loops modulo
conformal equivalence.  In this appendix, for completeness, we shall discuss how a conformal 
transformation would act on the couplings and the contour of our loops. We find useful first to investigate 
conformal transformations on the relevant Killing spinors and then to extend the analysis on the scalar couplings. 

\subsection{Conformal transformations on Killing spinors}
The simplest way to construct these transformations  is to view  the couple $(\epsilon_{s},\epsilon_{c})$
as a positive chiral spinor in the  spinor representation  of  the ten dimensional conformal group $SO(11,1)$.
This  representation is realized in terms of the $64\times 64$ Dirac matrices
\be
\label{sino}
\sigma_{A}\sigma_{B}+\sigma_{B}\sigma_{A}=2\eta_{AB}\mathds{1},
\ee
where 
$
\eta_{AB}=\mathrm{diag}(\underset{11}{1,\dots,1},-1).
$
The chiral  representation for the  $\sigma_{A}$  can be given in  terms of the following  block-antidiagonal matrices
\be
\sigma^{a}=\begin{pmatrix}0 & \gamma^{a}\\
\gamma^{a}& 0\end{pmatrix},\  \  \  (a=1,\dots,10) \ \ \ 
\sigma^{11}=\begin{pmatrix}0 & \gamma^{11}\\
\gamma^{11}& 0\end{pmatrix}\  \  \ 
\sigma^{12}=\begin{pmatrix}0 & \mathds{1}\\
-\mathds{1}& 0\end{pmatrix},
\ee
where $\gamma^{a}$ are the ten dimensional Euclidean Dirac matrices. Consider now a spinor  of positive chirality
\be
\sigma^{13}\begin{pmatrix}\epsilon\\ 0\end{pmatrix}=\begin{pmatrix}\epsilon\\ 0\end{pmatrix}.
\ee
The  action of the generators of $SO(11,1)$  on this spinor can be rewritten in ten dimensional language
as follows
\be
\label{sino1}
\begin{split}
M^{ab}\epsilon=\frac{1}{2}\gamma^{ab}\epsilon\  \ &\ \ (a,b=1,\dots,11) \ \ \ \ 
M^{12, a}\epsilon=\frac{1}{2}\gamma^{a}\epsilon\  \ \ \ (a=1,\dots,11),
\end{split}
\ee
where $\epsilon$ is now viewed as a 32 component Dirac spinor in  ten dimensions. There is an obvious  embedding of the conformal subgroup $SO(5,1)$ in the representation \eqref{sino1}.  It is obtained  by setting
\be
\begin{split}&M^{\mu\nu}\epsilon=\frac{1}{2}\gamma^{\mu\nu}\epsilon\  \ \ \ 
\Delta\epsilon=M^{12,11}\epsilon=\frac{1}{2}\gamma^{11}\epsilon\\
&P_{\mu}\epsilon=(M^{12,\mu} +M^{11,\mu})\epsilon=\frac{1}{2}\gamma^{\mu}\epsilon+\frac{1}{2}\gamma^{11,\mu}\epsilon=\frac{1}{2}\gamma^{\mu}(1-\gamma^{11})\epsilon\\
&K_{\mu}\epsilon=(M^{12,\mu} -M^{11,\mu})\epsilon=\frac{1}{2}\gamma^{\mu}\epsilon-\frac{1}{2}\gamma^{11,\mu}\epsilon=\frac{1}{2}\gamma^{\mu}(1+\gamma^{11})\epsilon,
\end{split}
\ee
with $\mu=1,\dots,4$.
We can also embed the R-symmetry group $SO(6)$ of $\mathcal{N}=4$ SYM  by choosing 
\be
R^{i-4,j-4}\epsilon=M^{ij}\epsilon=\frac{1}{2} \gamma^{ij}\epsilon,\ \ \ \  (i,j=5,\dots,10).
\ee
The original couple of ten dimensional chiral spinors $(\epsilon_{s},\epsilon_{c})$ is recovered by 
decomposing  the spinor $\epsilon$  in eigenstates of $\gamma^{11}$. We shall write
\be
\epsilon=\epsilon_{s}+ \epsilon_{c}\ \ \   \mathrm{with}\ \  \  \gamma^{11}\epsilon_{s}=\epsilon_{s} \ \ \ 
\mathrm{and}\ \ \ \ 
 \gamma^{11}\epsilon_{c}=-\epsilon_{c}.
\ee
The action of $P^{\mu}$, $K^{\mu}$ and $\Delta$ in terms of  $\epsilon_{s}$ and $\epsilon_{c}$ can be also rewrite as follows
\be
\label{sino2}
\begin{split}
&\Delta\epsilon_{s}=\frac{1}{2}\gamma^{11}\epsilon_{s}=\frac{\epsilon_{s}}{2}
\ \ \ \ 
\Delta\epsilon_{c}=\frac{1}{2}\gamma^{11}\epsilon_{s}=-\frac{\epsilon_{c}}{2}\\
&P_{\mu}\epsilon=\frac{1}{2}\gamma^{\mu}(1-\gamma^{11})\epsilon=\gamma_{\mu}\epsilon_{c},\\
&K_{\mu}\epsilon=\frac{1}{2}\gamma^{\mu}(1+\gamma^{11})\epsilon=\gamma_{\mu}\epsilon_{s}.
\end{split}
\ee
Exploiting \eqref{sino2} we can easily compute the action of a finite translation or of a finite special conformal transformation on the spinor $\epsilon$.  For a translation, we obtain
\be
\exp\left(v^{\mu} P_{\mu}\right)\epsilon=\epsilon+v^{\mu}\gamma_{\mu}\epsilon_{c}+\sum_{n=2}^{\infty}\frac{(-1)^{n}}{n!} (v^{\mu} P_{\mu})^{n}\epsilon=\epsilon+v^{\mu}\gamma_{\mu}\epsilon_{c},
\ee
 since 
 \be
 (v^{\alpha} P_{\alpha})^{2}\epsilon=\frac{1}{4}v^{\alpha}v^{\beta}\gamma_{\alpha}(1+\gamma^{11})
 \gamma_{\beta}(1+\gamma^{11})=0.
 \ee
 In other words under a {{}translation} the spinors $\epsilon_{s}$ and $\epsilon_{c}$ transform as follows
 \be
 {{}
 \epsilon_{c}\mapsto\epsilon_{c}\ \ \ \  \epsilon_{s}\mapsto \epsilon_{s}+v^{\mu}\gamma_{\mu}\epsilon_{c}.}
 \ee
 For a special conformal transformation, we find instead
 \be
 \exp\left(v^{\mu} K_{\mu}\right)\epsilon=\epsilon+v^{\mu}\gamma_{\mu}\epsilon_{s}+\sum_{n=2}^{\infty}\frac{(-1)^{n}}{n!} (v^{\mu} K_{\mu})^{n}\epsilon=\epsilon+v^{\mu}\gamma_{\mu}\epsilon_{s} \ee
 since  $(v^{\alpha} K_{\alpha})^{n}$ also vanishes for $n\ge 2$. In terms of the spinors $\epsilon_{s}$ and $\epsilon_{c}$, we have the following transformation
 \be
 \epsilon_{s}\mapsto\epsilon_{s}\ \ \ \  \epsilon_{c}\mapsto \epsilon_{c}+v^{\mu}\gamma_{\mu}\epsilon_{s}.
 \ee
The {dilation} instead yields
\be
{
\epsilon_{s}\mapsto e^{\frac{\lambda}{2}}\epsilon_{s}\ \ \ \  \epsilon_{c}\mapsto e^{-\frac{\lambda}{2}}\epsilon_{c}.}
\ee
The action of rotations and $SO(6)$ $R-$symmetry is instead the obvious one.

\subsection{Conformal transformations and contours}
In the following we shall illustrate how conformal transformations reflect on the set of parameters which defines
the Wilson loop.
\noindent
\paragraph{\sc Special conformal transformations:}
The  special conformal transformation  defined by the vector $v^{\mu}$ maps  $(\epsilon_{s},\epsilon_{c})$ to $(\epsilon_{s}^{\prime}, \epsilon_{c}^{\prime})=(\epsilon_{s}, \epsilon_{c}+v^{\mu}\Gamma_{\mu}\epsilon_{s})$.  The new parameters for the circuit are then given
\begin{align}
 a^{\prime\mu}=&\epsilon^{\prime}_{s}\Gamma^{\mu}\epsilon^{\prime}_{s}=\epsilon_{s}\Gamma^{\mu}\epsilon_{s}=a^{\mu}\nonumber\\
\lambda^{\prime}=&2\epsilon^{\prime}_{s}\epsilon_{c}^{\prime}=2\epsilon_{s}(\epsilon_{c}+v^{\alpha}\Gamma_{\alpha}\epsilon_{s})=\lambda+2(v\cdot a)\\
\hat \Omega^{\prime\mu}_{\,\,\, \nu}=&\epsilon_{s}^{\prime}\Gamma^{\mu}_{\,\, \nu}\epsilon_{c}^{\prime}=2\epsilon_{s}\Gamma^{\mu}_{\,\, \nu}(\epsilon_{c}+v^{\alpha}\Gamma_{\alpha}\epsilon_{s})=\Omega^{\mu}_{\,\, \nu}+2\epsilon_{s}\Gamma^{\mu}_{\,\, \nu}\Gamma_{\alpha}\epsilon_{s}v^{\alpha}=\Omega^{\mu}_{\,\, \nu}+ 2a^{\mu} v_{\nu}-2v^{\mu} a_{\nu},\nonumber
\end{align}
since $\epsilon_{s}\Gamma_{\lambda\mu\nu}\epsilon_{s}=0$. The parameter $b^{\prime\mu}$ is instead 
given by
\be
\begin{split}
\hat b^{\mu}=&\epsilon^{\prime}_{c}\Gamma^{\mu}\epsilon^{\prime}_{c}=
(\epsilon_{c}+v^{\beta}\epsilon_{s}\Gamma_{\beta})\Gamma^{\mu}(\epsilon_{c}+v^{\alpha}\Gamma_{\alpha}\epsilon_{s})=
\epsilon_{c}\Gamma^{\mu}\epsilon_{c}+2\epsilon_{s}\epsilon_{c} v^{\mu}-
2\epsilon_{s}\Gamma^{\mu}_{\,\, \alpha}\epsilon_{c} v^{\alpha}+\\
&+v^{\alpha}v^{\beta}\epsilon_{s}\Gamma_{\beta}\Gamma^{\mu}\Gamma_{\alpha}\epsilon_{s}=\epsilon_{c}\Gamma^{\mu}\epsilon_{c}+2\epsilon_{s}\epsilon_{c} v^{\mu}-
2\epsilon_{s}\Gamma^{\mu}_{\,\, \alpha}\epsilon_{c} v^{\alpha}+v^{\alpha}v^{\beta}\epsilon_{s}\Gamma_{\beta}\Gamma^{\mu}\Gamma_{\alpha}\epsilon_{s}=\\
=&b^{\mu}+\lambda v^{\mu}-\Omega^{\mu}_{\,\, \nu} v^{\nu} +2 v^{\mu} (a\cdot v)-v^{2} a^{\nu}.
\end{split}
\ee
The new circuit $y^{\mu}(s)$  is obviously obtained from the original one through the conformal transformation generated by the vector $v^{\mu}$
\be
\label{piop}
y^{\mu}=\frac{x^{\mu}-v^{\mu} x^{2}}{1-2(v\cdot x)+v^{2} x^{2}}.
\ee
The transformed couplings are instead given by
\be
\begin{split}
M^{\prime a}_{\,\, \mu}=&\epsilon_{s}^{\prime}\Gamma^{a}\tilde\Gamma_{\mu}\epsilon_{c}^{\prime}=
\epsilon_{s}\Gamma^{a}\tilde\Gamma_{\mu}\epsilon_{c}+v^{\alpha}\epsilon_{s}\Gamma^{a}\tilde\Gamma_{\mu}\Gamma_{\alpha}\epsilon_{s}=M^{a}_{\,\, \mu}+ B^{a}_{0} v_{\mu}\\
B^{\prime a}_{0}=& \epsilon_{s}^{\prime}\Gamma^{a}\epsilon^{\prime}_{s}=\epsilon_{s}\Gamma^{a}\epsilon_{s}=B^{a}_{0}\\
B^{\prime a}_{1}=& \epsilon_{c}^{\prime}\tilde\Gamma^{a}\epsilon^{\prime}_{c}=(\epsilon_{c}+v^{\beta}\epsilon_{s}\Gamma_{\beta})\tilde\Gamma^{a}(\epsilon_{c}+v^{\alpha}\Gamma_{\alpha}\epsilon_{s})=\\
=&\epsilon_{c}\tilde \Gamma^{a}\epsilon_{c}+2 v^{\beta}\epsilon_{s}\Gamma_{\beta}\tilde\Gamma^{a}\epsilon_{c}+v^{\beta} v^{\alpha} \epsilon_{s}\Gamma_{\beta}\tilde\Gamma^{a}\Gamma_{\alpha}\epsilon_{s}=B_{1}^{a}-2 v^{\beta} M^{a}_{\  \ \beta}-v^{2} B_{0}^{a}.
\end{split}
\ee
\paragraph{\sc Translations:} If we perform the translation defined by the vector $v^{\mu}$, the couple $(\epsilon_{s},\epsilon_{c})$ is mapped  to $(\epsilon_{s}^{\prime},
\epsilon_{c}^{\prime})=(\epsilon_{s}+v^{\mu}\Gamma_{\mu}\epsilon_{c}, \epsilon_{c})$, while the parameters   $(a^{\prime},\lambda^{\prime},
\Omega^{\prime},b^{\prime})$ become
\be
\begin{split}
a^{\prime\mu}=&\epsilon^{\prime}_{s}\Gamma^{\mu}\epsilon^{\prime}_{s}=
(\epsilon_{s}+v^{\beta} \epsilon_{c}\Gamma_{\beta})\Gamma^{\mu}(\epsilon_{s}+v^{\alpha}\Gamma_{\alpha} \epsilon_{c})=\\
=&
a^{\mu}+\lambda v^{\mu}+\Omega^{\mu}_{\,\, \nu} v^{\nu} +2 v^{\mu} (b\cdot v)-v^{2} b^{\nu}.
\\
\lambda^{\prime}=&2(\epsilon_{s}+u^{\alpha}\epsilon_{c}\Gamma_{\alpha})\epsilon_{c}=\lambda+2
(b\cdot v).
\\
\Omega^{\prime\mu}_{\,\,\, \nu}=&2\epsilon^{\prime}_{s}\epsilon_{c}^{\prime}=2 (\epsilon_{s}+v^{\alpha}\epsilon_{c}\Gamma_{\alpha})\Gamma^{\mu}_{\,\, \nu}\epsilon_{c}=
\Omega^{\mu}_{\,\, \nu}+2 v^{\mu} b_{\nu}-2 b^{\mu} v_{\nu}\\
b^{\prime\mu}=&\epsilon^{\prime}_{c}\Gamma^{\mu}\epsilon^{\prime}_{c}=\epsilon_{c}\Gamma^{\mu}\epsilon_{c}=b^{\mu}.
\end{split}
\ee
The new contour is obviously $y^{\mu}=x^{\mu}+v^{\mu}$, while the scalar couplings are 
\be
\begin{split}
M^{\prime a}_{\,\,\, \mu}=&\epsilon_{s}^{\prime}\Gamma^{a}\tilde\Gamma_{\mu}\epsilon_{c}^{\prime}=
\epsilon_{s}\Gamma^{a}\tilde\Gamma_{\mu}\epsilon_{c}+v^{\alpha}\epsilon_{c}
\tilde\Gamma_{\alpha}\Gamma^{a}\tilde\Gamma_{\mu}\epsilon_{c}=M^{a}_{\,\, \mu}- B^{a}_{1} v_{\mu}\\
B^{\prime a}_{1}=& \epsilon_{c}^{\prime}\tilde\Gamma^{a}\epsilon^{\prime}_{c}=\epsilon_{c}\tilde\Gamma^{a}\epsilon_{c}=B^{a}_{1}\\
B^{\prime a}_{0}=& \epsilon_{s}^{\prime}\Gamma^{a}\epsilon^{\prime}_{s}=(\epsilon_{s}+v^{\beta}\epsilon_{c}\tilde\Gamma_{\beta})\Gamma^{a}(\epsilon_{s}+v^{\alpha}\tilde\Gamma_{\alpha}\epsilon_{c})=\\
=&\epsilon_{c}\tilde \Gamma^{a}\epsilon_{c}+2 v^{\beta}\epsilon_{s}\Gamma^{a}\tilde\Gamma_{\beta}\epsilon_{c}+v^{\beta} v^{\alpha} \epsilon_{c}\tilde\Gamma_{\beta}\Gamma^{a}\tilde\Gamma_{\alpha}\epsilon_{c}=B_{0}^{a}+2 v^{\beta} M^{a}_{\,\, \beta}-v^{2} B_{1}^{a}.
\end{split}
\ee\paragraph{\sc Dilations:} When considering a dilation  the couple 
$(\epsilon_{s},\epsilon_{c})\mapsto(\epsilon_{s}^{\prime},
\epsilon_{c}^{\prime})=(e^{\rho/2}\epsilon_{s}, e^{-\rho/2}\epsilon_{c})$  and
$(a,\lambda,\Omega,b)\mapsto (e^{\rho}a,\lambda,\Omega, e^{-\rho}b)$. The new circuit is a
constant rescaling of the original one: $y=e^{\rho} x$. Finally the couplings are almost unchanged
\be
B_{0}^{\prime a}=e^{\rho} B_{0}^{a},\  \ \ \ B_{1}^{\prime a}=e^{-\rho} B_{1}^{a}\  \  \ \ \
\mathrm{and}\  \ \ \ \  M^{\prime a}_{\ \ \mu}=M^{a}_{\ \ \mu}.
\ee
We shall not discuss in details 
Lorentz rotations and $R-$symmetry transformations since they are realized on the circuit and on
the  couplings   in the obvious way.

\end{document}